\documentclass[fleqn,usenatbib]{mnras}

\usepackage{newtxtext,newtxmath}

\usepackage[T1]{fontenc}

\DeclareRobustCommand{\VAN}[3]{#2}
\let\VANthebibliography\thebibliography
\def\thebibliography{\DeclareRobustCommand{\VAN}[3]{##3}\VANthebibliography}

\usepackage{subfig}
\usepackage{tablefootnote}

\usepackage{graphicx}	
\usepackage{amsmath}	
\usepackage{siunitx}
\usepackage{multirow}
\usepackage{multicol}
\usepackage{hyperref}

\usepackage{diagbox}

\title[WAVES: Star-galaxy separation using UMAP]{Wide Area VISTA Extra-galactic Survey (WAVES): Unsupervised star-galaxy separation on the WAVES-Wide photometric input catalogue using UMAP and \textsc{hdbscan}}

\author[T. L. Cook et al.]{Todd L. Cook,$^{1}$\thanks{E-mail: t.cook@sussex.ac.uk}
Behnood Bandi,$^{1}$
Sam Philipsborn,$^{1}$
Jon Loveday,$^{1}$
Sabine Bellstedt,$^{2}$
Simon P. Driver,$^{2}$ \newauthor
Aaron S.G. Robotham,$^{2}$
Maciej Bilicki,$^{3}$
Gursharanjit Kaur,$^{3}$
Elmo Tempel,$^{4,5}$
Ivan Baldry, $^{6}$  \newauthor
Daniel Gruen, $^{7,8}$
Marcella Longhetti,$^{9}$
Angela Iovino,$^{9}$
Benne W. Holwerda, $^{10}$ 
and Ricardo Demarco$^{11}$
\\
$^{1}$Astronomy Centre, University of Sussex, Falmer, Brighton BN1 9QH, UK \\
$^{2}$International Centre for Radio Astronomy Research (ICRAR), M468, University of Western Australia, Crawley, WA 6009, Australia \\
$^{3}$Center for Theoretical Physics, Polish Academy of Sciences, al. Lotników 32/46, 02-668 Warsaw, Poland \\
$^{4}$Tartu Observatory, University of Tartu, Observatooriumi 1, 61602 Tõravere, Estonia \\
$^{5}$Estonian Academy of Sciences, Kohtu 6, 10130 Tallinn, Estonia \\
$^{6}$Astrophysics Research Institute, Liverpool John Moores University, IC2, Liverpool Science Park, 146 Brownlow Hill, Liverpool, L3 5RF, UK \\
$^{7}$University Observatory, Faculty of Physics, Ludwig-Maximilians-Universit\"at M\"unchen, Scheinerstr. 1, 81679 Munich, Germany \\
$^{8}$Excellence Cluster ORIGINS, Boltzmannstr. 2, 85748 Garching, Germany \\
$^{9}$INAF - Osservatorio Astronomico di Brera, via Brera 28, 20121 Milano - Italy \\
$^{10}$Department of Physics and Astronomy, University of Louisville, Natural Science Building 102, Louisville, KY 40292, USA \\
$^{11}$Institute of Astrophysics, Facultad de Ciencias Exactas, Universidad Andr\'es Bello, Sede Concepci\'on, Talcahuano, Chile \\
}

\date{Accepted 2024 October 15. Received 2024 September 2; in original form 2024 May 30}

\pubyear{2024}

\begin{document}
\label{firstpage}
\pagerange{\pageref{firstpage}--\pageref{lastpage}}
\maketitle

\begin{abstract}
Star-galaxy separation is a crucial step in creating target catalogues for extragalactic spectroscopic surveys. A classifier biased towards inclusivity risks including high numbers of stars, wasting fibre hours, while a more conservative classifier might overlook galaxies, compromising completeness and hence survey objectives. To avoid bias introduced by a training set in supervised methods, we employ an unsupervised machine learning approach. Using photometry from the Wide Area VISTA Extragalactic Survey (WAVES)-Wide catalogue comprising 9-band $u \ \mbox{-} \ K_s$ data, we create a feature space with colours, fluxes, and apparent size information extracted by \textsc{ProFound}. We apply the non-linear dimensionality reduction method UMAP (Uniform Manifold Approximation and Projection) combined with the classifier \textsc{hdbscan} to classify stars and galaxies. Our method is verified against a baseline colour and morphological method using a truth catalogue from Gaia, SDSS, GAMA, and DESI. We correctly identify 99.75\% of galaxies within the AB magnitude limit of  $Z=21.2$, with an F1 score of $0.9971 \pm 0.0018$ across the entire ground truth sample, compared to $0.9879 \pm 0.0088$ from the baseline method. Our method's higher purity ($0.9967 \pm 0.0021$) compared to the baseline ($0.9795 \pm 0.0172$) increases efficiency, identifying 11\% fewer galaxy or ambiguous sources, saving approximately 70,000 fibre hours on the 4MOST instrument. We achieve reliable classification statistics for challenging sources including quasars, compact galaxies, and low surface brightness galaxies, retrieving 92.7\%, 84.6\%, and 99.5\% of them respectively. Angular clustering analysis validates our classifications, showing consistency with expected galaxy clustering, regardless of the baseline classification.

\end{abstract}

\begin{keywords}
methods:data analysis --- surveys --- catalogues --- galaxies:photometry --- large-scale structure of Universe
\end{keywords}



\section{Introduction}

The classification of astronomical objects through their imaging is a key tool for astronomy. Spectroscopic surveys such as SDSS\footnote{\href{https://www.sdss.org/}{Sloan Digital Sky Survey}}, GAMA\footnote{\href{http://www.gama-survey.org/}{Galaxy and Mass Assembly survey}} and the upcoming surveys using the 4MOST\footnote{\href{https://www.4most.eu/cms/home/}{4-metre Multi Object Spectroscopic Telescope}} instrument (\citealp{york_sloan_2000,driver_galaxy_2011,de_jong_4most_2019} \ respectively) all require an input catalogue of selected targets. These targets are generated through the analysis of prior imaging, with the star-galaxy classification of the targets being a crucial step. For an extragalactic spectroscopic survey, a target catalogue based on a more liberal classifier will result in a high number of stars, leading to wasted fibre-hours. A more conservative classifier on the other hand will cause the omission of more galaxies in the spectroscopic observations.
Star-galaxy separation can be conducted by multiple different methods. Modern-day star-galaxy separation techniques can be split into colour, morphological and machine learning methods. The DEVILS survey \citep{davies_deep_2018} utilises NIR colours and surface brightness to filter stars from their target catalogue. The GAMA input catalogue (\citealp{baldry_galaxy_2010}) utilises SDSS imaging, and classifies sources into stars versus galaxies using a combination of profile fitting and colour separation. Morphological star-galaxy separation techniques are able to differentiate between extended sources (galaxies) and point sources. They cannot however differentiate between the nature of the point source, as they could be stars or quasi-stellar objects (QSOs). Morphological techniques such as \cite{slater_morphological_2020} and \cite{soumagnac_stargalaxy_2015} show promise but show limited reliability at faint magnitudes. This is due to astronomical seeing `smearing' psf-sized point sources such as stars and quasars, which is apparent even in space-based imaging (\citealp{holwerda_cosmic_2024}).

For decades, star-galaxy separation has been seen as an exemplary classification task for supervised machine learning (\citealp{odewahn_star-galaxy_1993,weir_automated_1995,bertin_sextractor_1996,bailer-jones_quasar_2019,clarke_identifying_2020,baqui_minijpas_2021}), with numerous models being used (neural networks, random forest, support vector machines etc.). However, supervised machine learning requires prior training data, with classifications being most effective when the training data is abundant and representative of the test data. With that said, there are ways of retrieving unbiased results for classifications or parameter estimation given a small amount of training data (e.g. using active learning; \citealp{lochner_astronomaly_2021,stevens_astronomical_2021}) or using a biased training set (e.g. \citealp{gruen_selection_2017}).

We utilise unsupervised machine learning for our star-galaxy separation, without the use of any training data. Unsupervised machine learning (including dimensionality reduction and clustering) has the advantage of making use of the entire given catalogue to find underlying clusters.  It also has the advantage of not being biased by a given training set, and is instead only dependent on the characteristics of the sample it is given. In this case, these include the depth of the survey, the elimination of artefacts etc. Instead of training data, we use data in which stars and galaxies are identified as such to high significance (which we term ``ground truth'' henceforth) purely for the purpose of validation. As we are only classifying stars and galaxies, it is easy to assign star or galaxy labels to the two clusters identified. This is as opposed to finding several distinct populations of sources such as in \cite{siudek_vimos_2018}.

Unsupervised machine learning has been used in astronomy before (A comprehensive recent review of unsupervised machine learning in astronomy can be found in \citealp{fotopoulou_review_2024}.) It has also been used specifically for star-galaxy separation before. \cite{logan_unsupervised_2020} also use \textsc{hdbscan} for star-galaxy-QSO separation with a different preprocessing stage. Instead of performing star-galaxy-QSO classification, we focus solely on distinguishing between stars and galaxies. This decision is based on WAVES' avoidance of quasars in their target selection process. If quasars are mistakenly classified as stars, they will not be in the target catalogue. However, if they are correctly classified as galaxies, they should be filtered out by WAVES' photometric redshift selection criteria and will not be part of the part of the target catalogue regardless. We investigate this in Section~\ref{quasars}.

In this work, we aim to provide star-galaxy separation for the input catalogue for the Wide region of the WAVES survey, part of the 4MOST suite of surveys. Our goal is to enable WAVES to achieve the 95\% completeness required for its science goals with as few fibre hours as possible. Firstly in Section~\ref{data}, we summarise the photometric catalogue of the WAVES target catalogue. We then describe how we compiled a sample of ground truth data we use for the validation of our method, and analyse how it compares to the overall target catalogue. In Section~\ref{baselinemethod}, we describe a baseline star-galaxy separation method currently used in WAVES as a point of comparison for our method. In Section~\ref{prepro}, we outline the preprocessing part of the method, including data cleaning, feature formation, feature scaling, and the dimensionality reduction which performs most of the `heavy lifting' of the method. In Section~\ref{hdbscan}, we explain the actual clustering method used to separate stars and galaxies in the reduced data. In Section~\ref{classification_performance}, we measure the fidelity of our method using the ground truth sample we compiled using confusion matrices and analysing F1 score as a function of magnitude. In Section~\ref{challening_galaxies}, we measure the effectiveness of our method at classifying galaxies that star-galaxy separation methods would find challenging. Finally in Section~\ref{clustering}, we examine the two-point angular correlation function of our sources.

\section{Data}
\label{data}

\begin{figure}
    \includegraphics[width=\columnwidth]{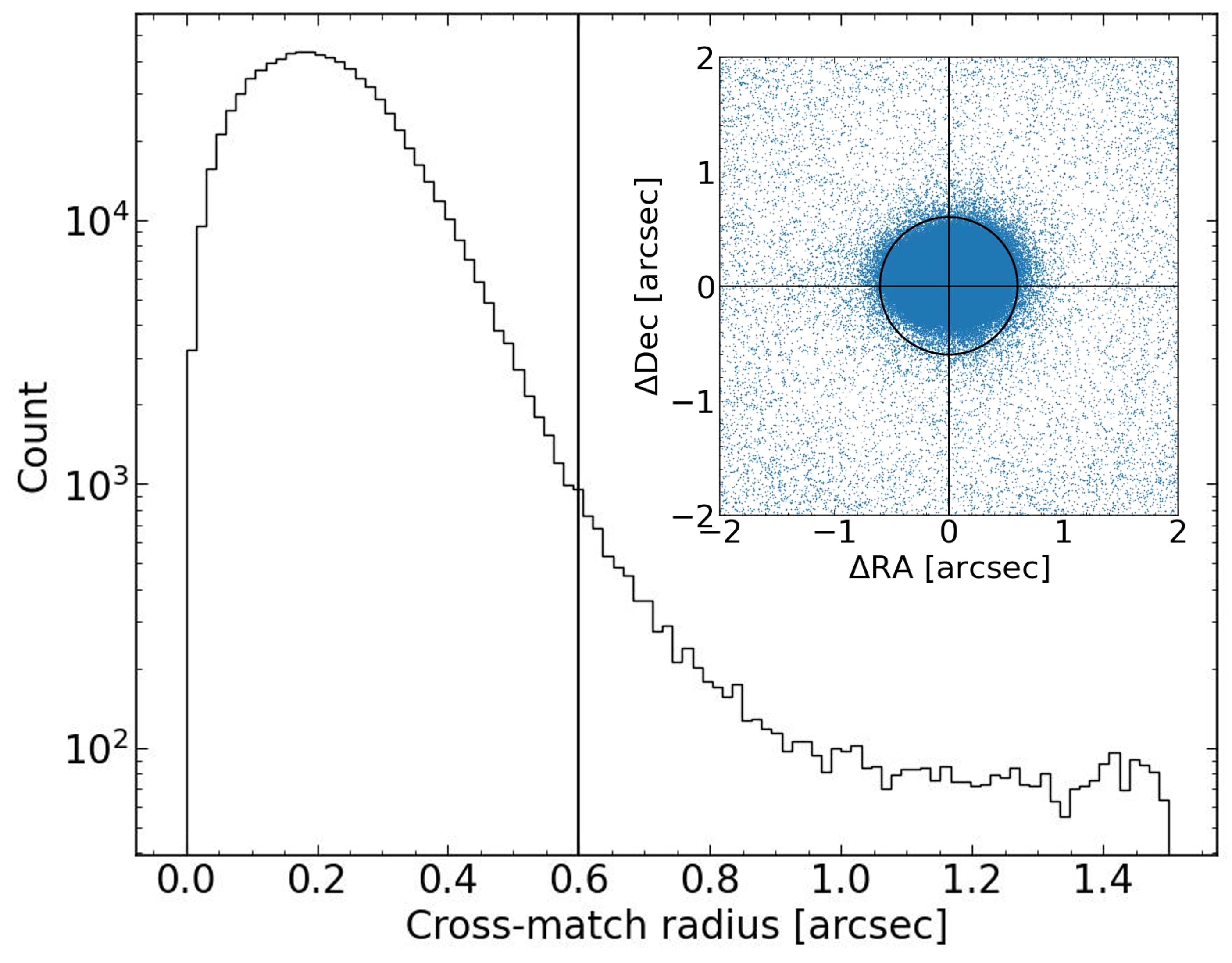}
    \caption{Histogram of the radial difference between sources in the WAVES-Wide photometric input catalogue and our ground truth catalogue, with a line showing our 0.6" cross-match radius. The upper-right panel shows the difference in RA (WAVES $-$ ground truth) and Dec, with a circle indicating the 0.6" cross-match radius.}
    \label{cross_match}
\end{figure}

The data utilised in this work is comprised of the multiband photometric data on which we classify objects, and a compiled sample of ground truth data which we use for verification. The photometric data are described in Section~\ref{photometric}, while the ground truth data are described in Section~\ref{groundtruth}.

\subsection{Photometric catalogue}
\label{photometric}
For our photometric data, we use the parent photometric catalogue of the upcoming WAVES-Wide survey (\citealp{driver_4most_2019}), a survey of the local Universe to be performed with the 4MOST spectroscopic instrument\footnote{4MOST is observing different types of targets (e.g. stars, galaxies, quasars) simultaneously as if they were one single survey (\citealp{tempel_optimized_2020,tempel_probabilistic_2020})}. WAVES will use approximately 1.75 million low resolution fibre hours providing spectra at $ R=4000 \ \mbox{-} \ 7000$ covering the wavelength range from $370 \ \mbox{-} \ \SI{950}{nm}$. WAVES-Wide covers an area of \SI{\sim 1170}{\mathrm{deg^2}} over its North and South regions, and contains \SI{\sim 14800000}{} sources within the $Z<21.2$ magnitude limit of the parent catalogue\footnote{The actual WAVES magnitude limit will likely be brighter in both the WAVES and Deep regions}. The North region lies on the equatorial plane and spans 157.25 to 225.0 degrees in Right Ascension. The South region sits at -30 degrees Declination and spans -30 to 52.5 degrees Right Ascension.

Catalogue construction will be fully be outlined in Bellstedt et al. (in prep.), and is derived from deep optical and NIR imaging from VST KiDS\footnote{\href{https://kids.strw.leidenuniv.nl/}{VLT Survey Telescope Kilo-Degree Survey}} (\citealp{de_jong_kilo-degree_2013,kuijken_fourth_2019}) (with the $u,g,r$ and $i$ bands) and the VISTA VIKING\footnote{\href{https://www.eso.org/public/teles-instr/paranal-observatory/surveytelescopes/vista/surveys/}{Visible and Infrared Survey Telescope for Astronomy VISTA Kilo-degree Infrared Galaxy}} survey \citep{edge_vista_2013} (with the $Z, Y, J, H$ and $K_s$ bands) respectively. We note that far- and mid-infrared photometry also provide useful information for star-galaxy separation (\citealp{kovacs_stargalaxy_2015,kurcz_towards_2016,krakowski_machine-learning_2016}). \cite{logan_unsupervised_2020} note that the use of $W1$ and $W2$ bands from WISE\footnote{\href{https://www.jpl.nasa.gov/missions/wide-field-infrared-survey-explorer-wise}{Wide-field Infrared Survey Explorer}} \citep{wright_wide-field_2010} in their unsupervised star-galaxy-quasar classifier are critical for their precision and accuracy. The Random Forest analysis of the features identifies $W1$ and $W2$ as the most important bands. However, \cite{nakoneczny_photometric_2021} using KiDS and VIKING can effectively classify quasars even without WISE data. We do not incorporate WISE data in this work because the corresponding observations are not deep enough compared to the KiDS and VIKING bands. After background subtraction, a large fraction of sources have negative WISE fluxes which are clearly unphysical, and cannot be converted to magnitudes. As demonstrated in further sections, we are confident this work has the capability to reliably classify faint targets without WISE data.

Source detection and characterization are carried out by the \textsc{ProFound} package (\citealp{robotham_ProFound_2018}). An inverse variance weighted stack of $r+i+Z+Y$ bands is utilised for initial source detection. \textsc{ProFound} works by detecting and maintaining the original isophote (or segment) for a source, which can vary in shape from regular to irregular. Then, segment dilation occurs to derive pseudo-total fluxes. This involves progressively adding layers of pixels surrounding each segment until the flux reaches convergence. Segments for each source are generated and large, fragmented sources (see Figure~11 of \citealp{bellstedt_galaxy_2020}) are manually regrouped after being flagged by the \textsc{ProFound} pipeline. The radius and ellipticity of each source is determined depending on the size and shape of each segment. The resulting flux and flux error are estimated for each band within each segment.

The Planck $\mathrm{E(B-V)}$ extinction map is applied to the sources (\citealp{planck_collaboration_planck_2013}), correcting their flux for Galactic dust absorption. Finally, stars brighter than a $G$-band magnitude of 16.0 are removed, and all sources within a radius of $10^{1.6-0.15G}$ arcminutes of these bright stars are masked out because their flux can affect the estimate of other sources' fluxes in the photometry, where $G$ is the Gaia $G$-band magnitude. This results in brighter stars having a larger exclusion radius. After star masking, we are left with \SI{14802032}{} sources within the $Z<21.2$ magnitude limit that need to be classified.

The WAVES target catalogue will also be limited by a photometric redshift upper limit of $z<0.2$ in the Wide fields and $z<0.8$ in the Deep fields\footnote{We do not address the Deep fields in this work, but our method is used for the Deep fields in the WAVES target selection}. Each source will have an associated photometric redshift, which are found by combining multiple photo-z methods (detailed in Bellstedt at al. in prep.) using the same KiDS + VIKING photometry. All star-galaxy separation conducted in this work is prior to the photo-z selection cuts.

\subsection{Ground truth catalogue}
\label{groundtruth}

\begin{figure*}
    \includegraphics[width=\textwidth]{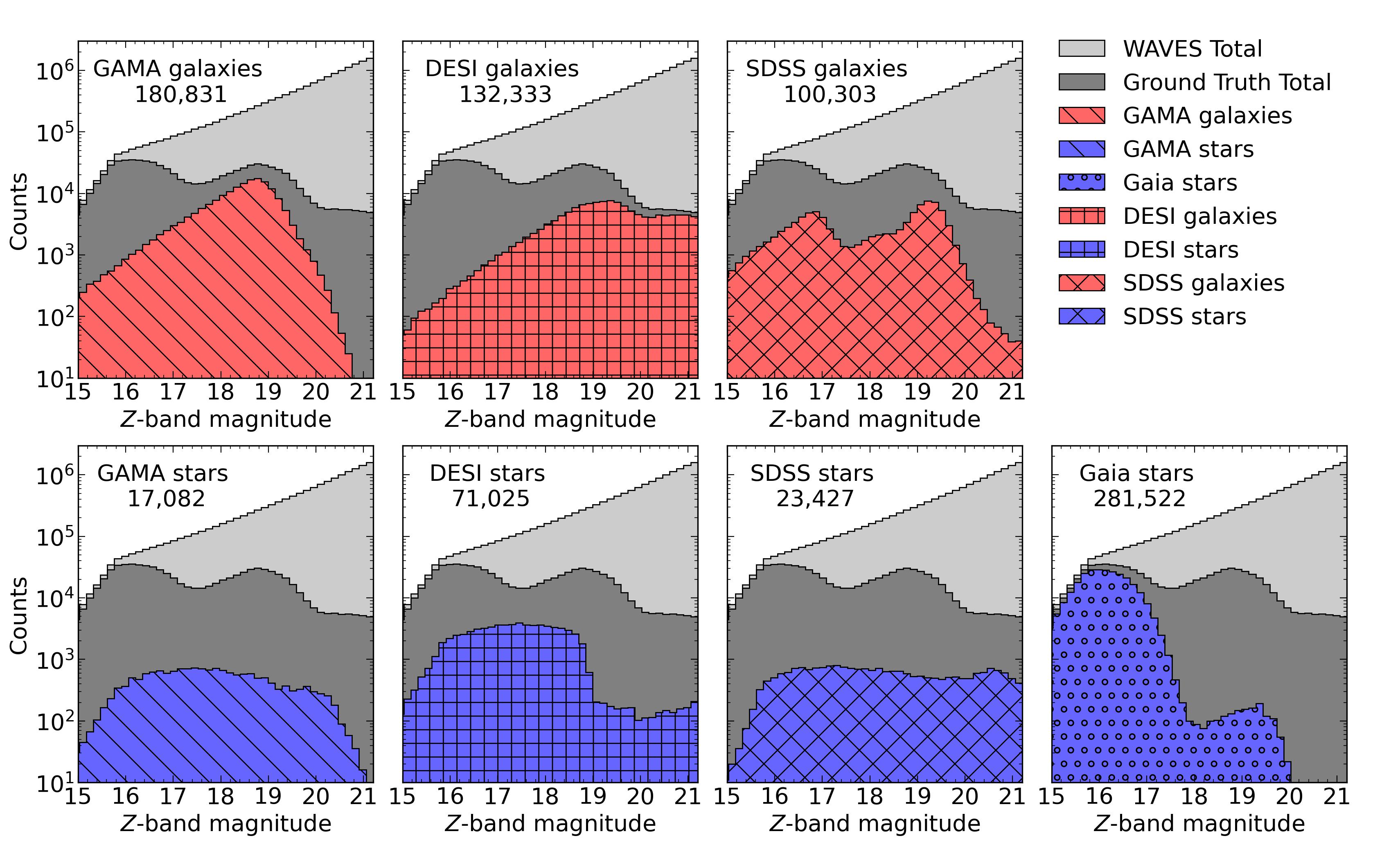}
    \caption{Histograms of the ground truth data sourced for this analysis compared to the total number of sources needed to be classified in the WAVES-Wide regions (light grey) as a function of their $Z$-band magnitude. Red and blue regions notate galaxies and stars respectively. The different hatchings notate the different sources of ground truth data (GAMA, SDSS, DESI and Gaia). The dark grey region indicates the total number of ground-truth sources. The ratio of ground truth data to the total number of WAVES sources drops off quickly at faint magnitudes. }
    \label{fig:z_mag_count}
\end{figure*}

To assess the performance of our classifier, we compile a catalogue of sources with known classification as a test set. As we cannot confidently infer ground truth classification from photometry, we rely on pre-existing spectroscopic data and Gaia parallax measurements in the WAVES-Wide fields.
When matching sources between the WAVES input catalogue and the ground truth data, we use a 0.6" cross-match radius. This can be visualised in Figure~\ref{cross_match}, which shows a histogram of the radial difference between sources in the WAVES-Wide photometric catalogue and sources in the ground truth catalogue. The upper right panel shows the differences in RA and Dec, showing a small systematic offset. We choose 0.6" as a conservative cross-match radius, preventing many potential spurious matches, as we prioritise purity in the ground truth dataset. For WAVES equatorial coordinates, we use \texttt{RAmax} and \texttt{Decmax} generated from the \textsc{ProFound} package, which correspond to the position of the pixel which contains the greatest flux in the segment.

\subsubsection{GAMA}
The GAMA survey is an extragalactic spectroscopic survey conducted with the AAOmega wide-field facility on the Anglo Australian Telescope. GAMA consists of four \SI{\sim 50}{deg^2} equatorial fields located at $\mathrm{2^h}$ (G02), $\mathrm{9^h}$ (G09), $\mathrm{12^h}$ (G12), $\mathrm{15^h}$ (G15) and one field at $\mathrm{23^h}$ and $\mathrm{-32.5^\circ}$ (G23). Two of these fields lie within the WAVES-Wide North equatorial region, and G23 lies within the WAVES-Wide South region. GAMA is a strictly extragalactic survey. However, due to the imperfect star-galaxy separation used to define the target catalogue, a number of stars were observed.  We use the fourth data release of GAMA (\citealp{driver_galaxy_2022}), which also utilises KiDS and VIKING photometry. We filter \texttt{gkvScienceCatv02}\footnote{\url{https://www.gama-survey.org/dr4/schema/}} by $\texttt{NQ} > 2$, meaning a reliable spectroscopic redshift was measured. After a cross-match with our sample, we retrieve \SI{180831} galaxies, \SI{17082} stars and 7 ambiguous sources that have been classified through GAMA's analysis of spectral features and redshift (\citealp{hopkins_galaxy_2013}). We do not apply any additional redshift cuts in distinguishing stars and galaxies. We discard the 7 ambiguous sources.

Due to the magnitude limit of GAMA ($r\lesssim 19.65$), the majority of the sample lies well below the $Z<21.2$ magnitude limit of this work. This can be seen in the first column of plots in Figure~\ref{fig:z_mag_count}. GAMA provides a plethora of galaxies up to $Z<19$, but the distribution drops off quickly, and no GAMA galaxies or stars lie at the magnitude limit of this work. GAMA does however have the advantage of being highly complete, and is not biased against any particular galaxy type.

\subsubsection{DESI}
For sources fainter than the magnitude limit of GAMA, we utilise the Early Data Release from the Dark Energy Spectroscopic Instrument (DESI; \citealp{desi_collaboration_early_2023}). DESI is mounted on the 4-m Mayall telescope at Kitt Peak National Observatory in Arizona and will take spectra of roughly 40 million galaxies and quasars over its 5 year programme  over an area of \SI{14000}{deg^2} (\citealp{desi_collaboration_overview_2022}). Whilst DESI is primarily targeting the Northern sky, its footprint does overlap with the equatorial WAVES-Wide North region. We find \SI{132333}{} galaxies and \SI{71025}{} stars cross-matched to sources in the WAVES-Wide catalogue from DESI's `Target Selection Validation' (SV1) and `One-Percent Survey' (SV3). We filter \texttt{zall-tilecumulative-edr-vac} by $\texttt{OBJTYPE=TGT}$ which excludes faulty and sky fibers, and remove any sources with problematic spectra using $\texttt{ZWARN==0}$. We also remove galaxies with a redshift lower than $z<0.0015$, and remove stars with a redshift higher than $z>0.0015$ to prevent contamination.

The DESI galaxy sample contains three tracers: bright galaxies (\citealp{hahn_desi_2023}) which are magnitude limited down to $r < 19.5$ and colour-selected in the magnitude range $19.5<r<20.175$, luminous red galaxies (\citealp{zhou_target_2023}) which are colour-selected down to $Z-$band fiber magnitude of  $Z<21.6$, and emission line galaxies (\citealp{raichoor_target_2023}) which are colour selected in the magnitude range $g>20$ and $g-$band fibre magnitude of  $g_\mathrm{fibre}<24.1$. The DESI Milky Way Survey (\citealp{prieto_preliminary_2020,cooper_overview_2023}) select Gaia colour-selected stars in the magnitude range $16 < r < 19$, and provide the majority of stars used for our sample. The magnitude distribution of DESI sources can be seen in the second column of Figure~\ref{fig:z_mag_count}. Visual inspection of DESI galaxies (\citealp{lan_desi_2023}) and quasars (\citealp{alexander_desi_2023}) reveal good classification of galaxies ($\sim99\%$ purity), and reasonable classification of quasars ($\sim86\%$ purity), although the spectroscopic classification pipeline for the EDR has been updated since. The relative depth of DESI is critical for the validation of the star-galaxy separation, as it provides spectra for sources all the way down to the magnitude limit of this analysis.

\subsubsection{SDSS}
We also utilise data from the 17th data release of SDSS (\citealp{abdurrouf_seventeenth_2022}) for our validation. SDSS has a plethora of spectroscopically confirmed galactic and extragalactic sources in the WAVES-Wide North equatorial region, with some reaching out to the magnitude limit of $Z<21.2$. We filter our SDSS sample by $\texttt{ZWARNING==0}$, implying that a reliable redshift measurement has been found. We are able to cross-match \SI{123730}{} sources with SDSS DR17. Of these sources, 41.6\% are galaxies from the SDSS main survey (\citealp{strauss_spectroscopic_2002}) and 52.0\% are galaxies from the Baryon Oscillation Spectroscopic Survey (BOSS; \citealp{dawson_baryon_2013}). This sample also contains stars from the stellar survey, the Sloan Extension for Galactic Understanding and Exploration (SEGUE; \citealp{yanny_segue_2009}), although the vast majority of faint stars come from the BOSS survey. In fact, of the stars that are sourced from SDSS, 93.4\% of those between the magnitudes $20.0<Z<21.2$ are misclassified quasars from BOSS (\citealp{ross_sdss-iii_2012}).

\subsubsection{Gaia}
Gaia is a space-based telescope, designed to create an all-sky optical map, mainly focusing on the stars in our galaxy (\citealp{Gaia_collaboration_Gaia_2016}) but also including extragalactic sources such as quasars (\citealp{Gaia_collaboration_Gaia_2023}). We make use of Gaia's third data release (\citealp{brown_Gaia_2021}) which contains astrometry and photometry for 1.8 billion sources. We use the Gaia archive query described in Appendix B of \cite{Gaia_collaboration_Gaia_2018} to extract a significant number of stars in the WAVES-Wide regions. This query utilises Gaia's parallax measurements and ensures each source has a $R$, $G$ and $B$ flux with a signal-to-noise ratio of at least 10. Ultimately we are able to generate a catalogue of \SI{281522}{} sources that we are confident are stars in the WAVES-Wide regions. The resulting magnitude distribution of these stars can be seen in the fourth plot on the bottom row of Figure~\ref{fig:z_mag_count}. As parallax measurements require a high signal-to-noise ratio, the stars we retrieve from Gaia are significantly biased towards brighter magnitudes.

\subsubsection{Compilation}

We find that \SI{102904}{} sources overlap across different surveys, most of these being between GAMA, DESI and SDSS observations in the WAVES North region. We find 218 sources with contradictory labels between surveys, which we remove. Overall, our ground-truth catalogue consists of \SI{367888}{} stars and \SI{335295}{} galaxies, with a ratio of stars to galaxies roughly what we expect from the WAVES catalogue at fainter magnitudes. Figure~\ref{fig:z_mag_count} shows the number counts of sources with ground truth labels compared to the total sources in the WAVES-Wide regions as a function of $Z$-band magnitude. This plot demonstrates the advantage of unsupervised over supervised machine learning in this problem. At the 21.2 $Z$-band magnitude limit, only 0.27\% of sources have a ground truth label, which provides a challenge regardless of the method of star-galaxy separation. Supervised machine learning methods are, however, known to be heavily biased by the training data they receive. Our ground-truth labels do not constitute a significant fraction of the total number of sources, as a result, we make no use of the ground-truth labels in the classification of objects, and use them purely for the purpose of verification.

\section{Baseline method}
\label{baselinemethod}

\begin{figure}
    \includegraphics[width=\columnwidth]{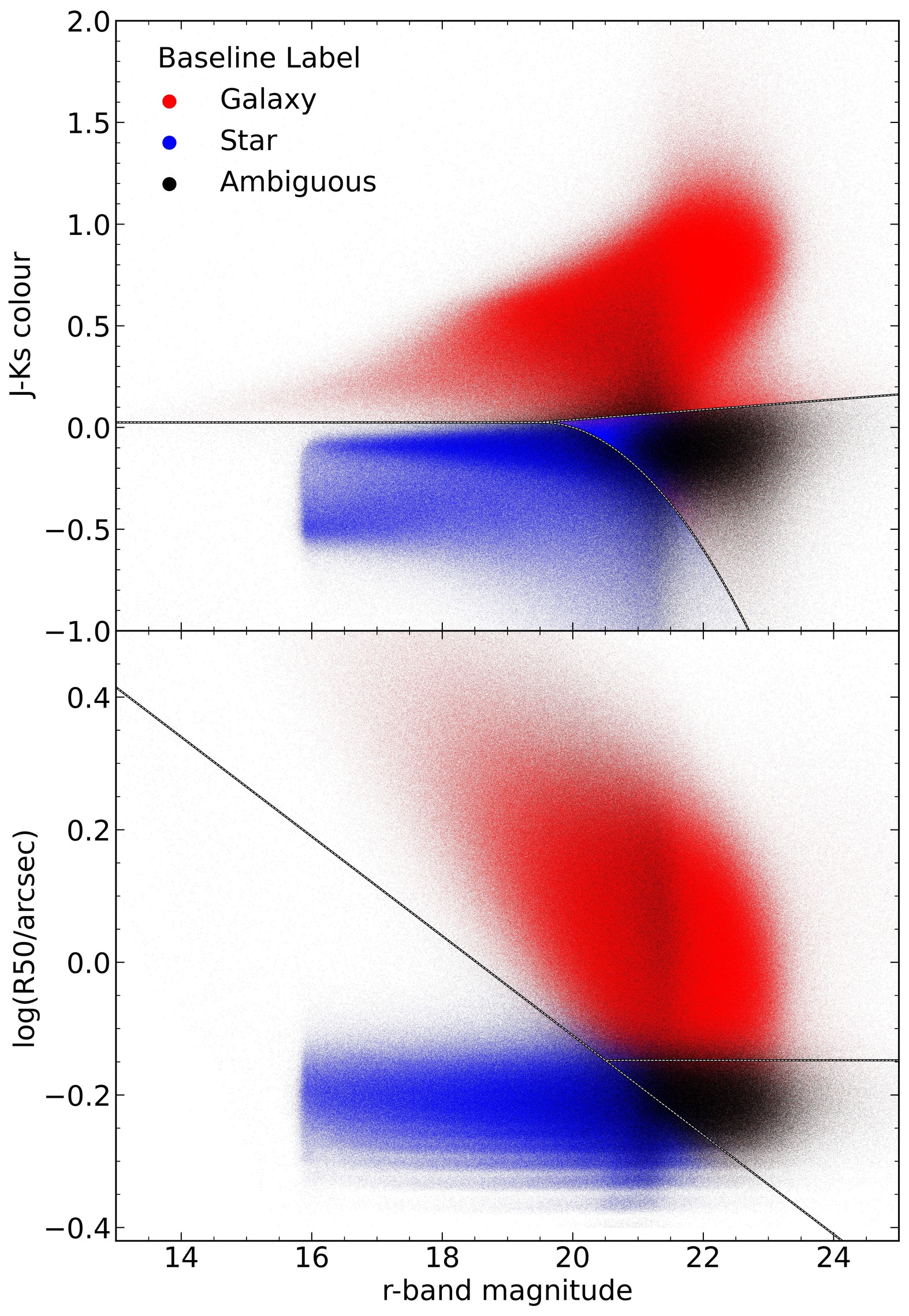}
    \caption{The baseline star-galaxy separation algorithm outlined in Section~\ref{baselinemethod} for all \SI{14802032}{} sources in the WAVES-Wide fields within the $Z<21.2$ magnitude limit. The upper panel shows $J-K_s$ colour as a function of $r$-band magnitude. The lower panel shows the log of the half-light radius, $R_{50}$ as a function of  $r$-band magnitude. The solid lines indicate the galaxy, star and ambiguous regions of the plots, with the colour indicating the final baseline classification. The sharp cut in sources brighter than $r = 16$ is due to the Gaia star mask, in which all stars brighter than a $G$-band magnitude of 16.0 are removed. The discretisation of the smallest $\log(R_{50})$ values is due to the discrete seeing values across tiles.}
    \label{fig:r_j-k_r50}
\end{figure}

One of the ways in which we assess the performance of our star-galaxy separation is by comparing against a baseline algorithm. We choose the classification algorithm used by GAMA DR4 (\citealp{driver_galaxy_2022}) outlined in Section~2.9.1 of \cite{bellstedt_galaxy_2020}, which uses very similar photometry. This algorithm utilises a combination of colour and size criteria.

The colours are derived from the `total' magnitudes in different bands. This involves adding the total flux within the segment of each source, estimated on the detection bands, and then converting the flux to magnitudes. This is different from `colour' magnitudes, which are derived using a fixed aperture across the multiple bands. For size information, the angular half-light radius $R_{50}$ is used. This is the radius in arcseconds that contains half of the detection band flux within the segment.

The classification method can be visualised in Figure~\ref{fig:r_j-k_r50}. Essentially, sources are plotted first in ($J-K_s$) colour versus $r$-band magnitude space and then in log $R_{50}$ versus $r$-band magnitude space. By doing this, stars and galaxies appear to occupy separate regions on the plots: galaxies tend to have larger ($J-K_s$) colours and increase in apparent radius as a function of brightness unlike stars. From this, lines are drawn to classify the sources into galaxy, star and ambiguous regions through the equations

\begin{align}
\begin{split}
    (J-K_s) = 0.025, & \qquad\text{if } r < 19.5\\
    (J-K_s) = 0.025+0.025(r-19.5), & \qquad\text{if }r > 19.5\\
    (J-K_s) = 0.025-0.1(r-19.5)^2, & \qquad\text{if }r > 19.5,
\end{split}
\end{align}
where the ambiguous region lies between the two lines beyond $r>19.5$ and:
\begin{align}
\label{baseline_equation}
\begin{split}
    \log(R_{50}) = \Gamma+0.05-0.075(r-20.5), & \quad\text{any }r\\
    \log(R_{50}) = \Gamma+0.05, & \quad\text{if }r > 20.5,
\end{split}
\end{align}
where $\Gamma$ is the median \textsc{log10seeing} value, the log of the seeing in arcseconds in the detection band ($r+i+Z+Y$). This value varies between -0.3 and -0.1 log(seeing/arcsec) depending on the tile.

Initially, all sources are labelled ambiguous. If a source is in the same region for both planes, then that region label is adopted. If a source is in differing star/galaxy regions on each plane, then it keeps its ambiguous label. And if it is ambiguous in one plane but star/galaxy in the other plane, then it adopts that class. One of the obvious drawbacks of this method is that it is likely to misclassify quasars, which are near point sources, especially if they have star-like colours.

\section{Preprocessing and dimensionality reduction}
\label{prepro}

The preprocessing stage of machine learning is critical, and plays a significant role in the final classification. In this work, the preprocessing and dimensionality reduction means that the final classification algorithm \textsc{hdbscan} has no trouble in separating the star and galaxy clusters in the lower-dimension space. This process improves the accuracy of the final classifier, and also reduces the computing resources necessary to make the classification.

\subsection{Data cleaning}

Firstly, all artefacts are removed from the catalogue. Artefacts are identified as sources with extremely unusual colours as outlined in Bellstedt et al. (in prep.). We then remove sources with any missing photometry from any band, as our dimensionality reduction method (UMAP) does not work with missing data. This equates to $1.07\%$ of the catalogue within the adopted magnitude limit of $Z < 21.2$. We then remove sources which have a negative flux after sky subtraction, as magnitudes cannot be calculated with negative flux. This removes a further $0.73\%$ of sources from the catalogue. Investigations were conducted using raw fluxes as the features instead of magnitudes, but this did not prove nearly as effective as using magnitudes.

One of the features used in star/galaxy separation is the KiDS $u$-band magnitude. However, a significant fraction, $10.57\%$, of sources have a negative $u$-band flux after sky subtraction. We found that the $u$-band does not significantly improve the quality of the star-galaxy separation overall, but does help to distinguish stars from quasars due to quasars' UV excess (\citealp{malkan_ultraviolet_1983}). We therefore run our classification algorithm twice, first with the $u$ band and then without, and combine the classifications, prioritising the label given by the run including the $u$-band. This means we only have to discard $1.80\%$ of sources in total.

We note the omission of sources with missing data is a drawback of this method, and the baseline method would be required for sources with missing data for the formation of a target catalogue. Alternatively, data imputation could be utilised such as in \cite{miller_preparing_2017}, although this is computationally expensive and requires further study. SED fitting could also be used to fill in missing data, as some photometric redshift template methods do not need all bands to function. This has the advantage of being physically motivated.

\subsection{Feature formation}
We input the magnitudes from all 9 available fluxes as features, using the total flux output from \textsc{ProFound}. We also use as all 36 possible colour combinations of the photometric bands using the colour flux output from \textsc{ProFound}. The colour flux is used to calculate colours as the segment is the same size in each band, unlike with total flux. Our colours will include some redundant information (e.g. $g-i$ colour is the same as ($g-r$) + ($r-i$)). However, we find that the dimensionality reduction method works best when all possible information is given to it, so that the most important features can be extracted. We also include as features the log of $R_{50}$, the effective half-light radius, and the axial ratio of each source. Finally, we include the log of the astronomical seeing of each source which varies from tile to tile. This leaves us with a total of 48 features for each source or 34 without $u$-band. \textsc{ProFound} does output flux errors however we do not incorporate them as features.

\subsection{Scaling}
We first use the \texttt{StandardScaler} package from \texttt{scikit-learn} (\citealp{pedregosa_scikit-learn_2011}) to appropriately scale our data. This involves scaling each feature to have a mean of 0 and unit variance (Z-score transformation), meaning that no feature is prioritised over another, thus avoiding bias.

\subsection{Dimensionality reduction using UMAP}

\begin{figure*}
    \includegraphics[width=\textwidth]{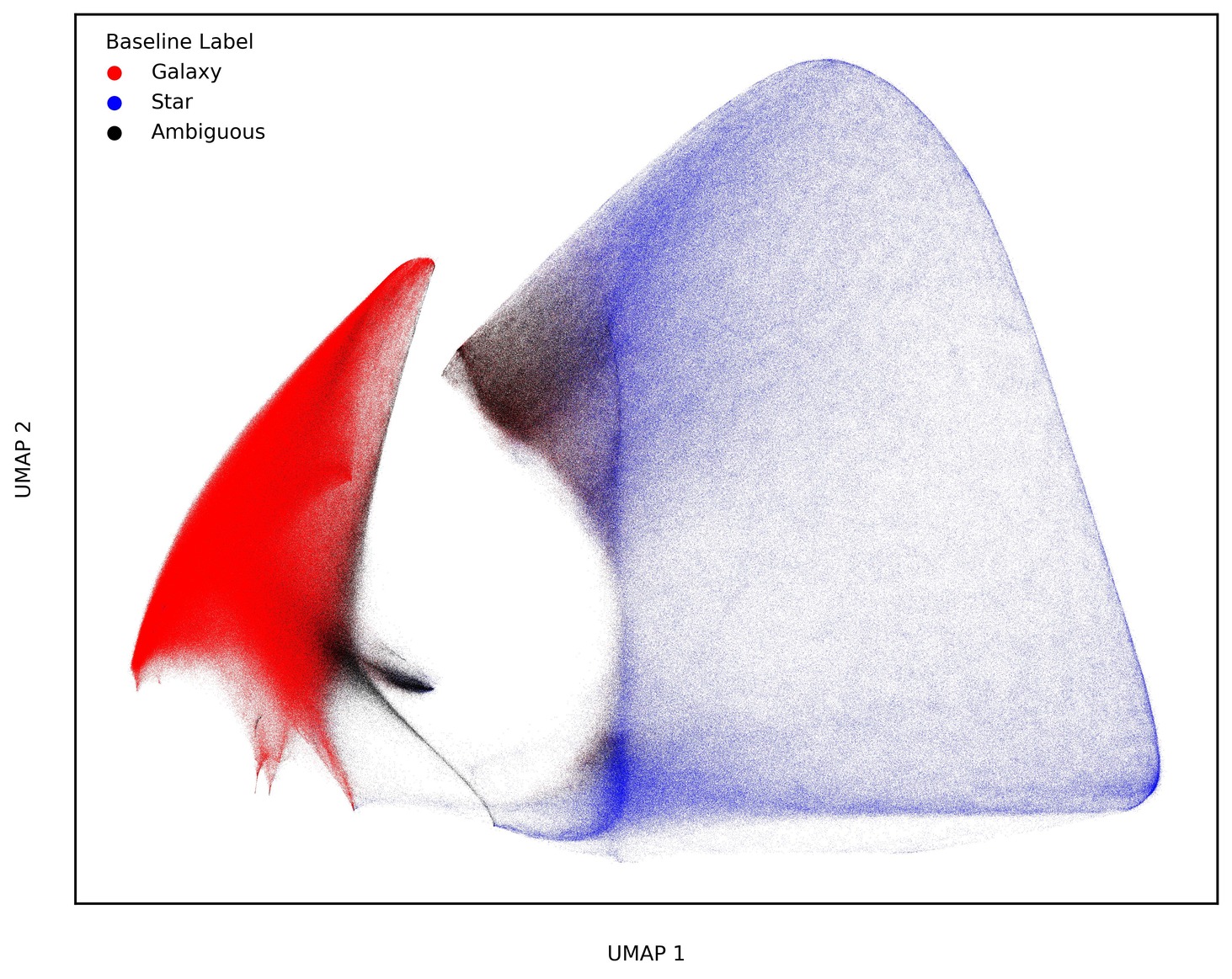}
    \caption{The first two UMAP projected dimensions generated from the \SI{6593731}{} sources of the WAVES-Wide South region within the $Z<21.2$ magnitude limit. The colours indicate the class label determined by the baseline method described in Section~\ref{baselinemethod}. Clearly, the sources have been separated into stars and galaxy `nodes' as indicated by separation in baseline class label. Ambiguous sources appear distributed in both nodes.}
    \label{fig:umap1_umap2_waves}
\end{figure*}

The majority of the `heavy lifting' of our star-galaxy separation is performed using Uniform Manifold Approximation and Projection (UMAP; \citealp{mcinnes_umap_2020}), a dimensionality reduction technique. We use dimensionality reduction due to the expensive cost of running a classifier on all 48 features, and we also find that using UMAP improves classification performance. UMAP works by first constructing a graph of nearest neighbours in high-dimensional space, where each data point is connected to its $n$~-nearest neighbours where $n$ is set by the user. Each vertex between points in this graph is weighted by the probability that these points are connected based on their distance, creating a `fuzzy' structure. This high-dimensional space is then projected to lower dimensional embeddings, where the layout of the graph is maintained as much as possible. UMAP is able to preserve both the global and local structure of the data. In the context of astronomical imaging, the global structure would be the separation between stars, galaxies, and quasars, and the local structure could be the properties of each source such as stellar classification, galaxy redshift and galaxy morphology.

UMAP works in a similar way to t-distributed Stochastic Neighbor Embedding (t-SNE) which has been implemented in astronomy research previously (\citealp{traven_galah_2017,anders_dissecting_2018, reis_detecting_2018,nakoneczny_photometric_2021,queiroz_starhorse_2023,guiglion_beyond_2024} etc.). However, UMAP has been demonstrated to be more scalable than t-SNE and better able to preserve the global structure of the data (\citealp{becht_dimensionality_2019, mcinnes_umap_2020}). UMAP is also advantageous to Principal Component Analysis (PCA, a review of which can be found in \citealp{jolliffe_principal_2016})  despite the latter's prolific use in data science, as UMAP is able to capture complex non-linear relationships within the data unlike PCA which is a linear dimensionality reduction technique. Self-organizing maps are also a good candidate for unsupervised star-galaxy separation. They have been used widely in astronomy for classification purposes such as object classification (\citealp{geach_unsupervised_2012}) and and identifying galaxy populations (\citealp{holwerda_galaxy_2022}), but also regression problems such as photometric redshifts (\citealp{masters_mapping_2015, wright_photometric_2020}).

Several star-galaxy separation methods have used UMAP for data visualisation and validation of their algorithm (\citealp{clarke_identifying_2020,stoppa_autosourceid-classifier_2023}), but fall short of using it for the main classification. Whilst there are some concerns that UMAP finds `mirages' of spurious local structure (\citealp{chari_specious_2023}), we are confident in its ability to find the global structure of our data in this context: distinguishing stars and galaxies, due to their differences in apparent features.

For our UMAP hyperparameters, we use 200 nearest neighbours and a minimum distance of 0. After running UMAP several times, these hyperparameters most often result in two clearly visible clusters in the UMAP feature space. We choose to reduce our 48-dimensional data down to 10 dimensions. This significantly reduces the computational power required for the classifier whilst maintaining some higher-dimensional correlations. Using a similar number of dimensions (e.g. 8,9,11,12) gives very similar results.

The first two projected dimensions of of UMAP applied to the \SI{6593731}{} sources of the WAVES-Wide South region can be seen in Figure~\ref{fig:umap1_umap2_waves}. We run UMAP on the North and South regions separately in case there are systematic differences between the two regions, and for computational reasons. The plot is coloured by each source's label given by the baseline algorithm described in Section~\ref{baselinemethod}. Clearly, sources appear to be clustered into two distinct nodes, and the baseline algorithm classification indicates that the left-hand node contains galaxies, and the right-hand node contains stars. Ambiguous sources appear to be distributed between both nodes, but are also densely populated in an area attached to the main galaxy node. These sources are primarily QSOs (explained in further sections). The sources `connecting' the two nodes are primarily blended sources with a mixture of flux from both galaxies and stars. This can happen when the source-finding algorithm mistakes multiple sources for a single source. In this case, the flux from foreground stars are contaminating the flux from background galaxies. We show all ten projected dimensions of UMAP in Figure~\ref{fig:UMAP_corner} in Appendix~\ref{appendix:corner}, visualising the entire 10-dimensional space which we use for clustering. Figure~\ref{fig:umap1_umap2_waves} is a positive indication of the fidelity of the algorithm, showing a general agreement between the UMAP embeddings of the data and the baseline algorithm.

\section{Classification with \textsc{hdbscan}}
\label{hdbscan}

\begin{figure*}
    \includegraphics[width=\textwidth]{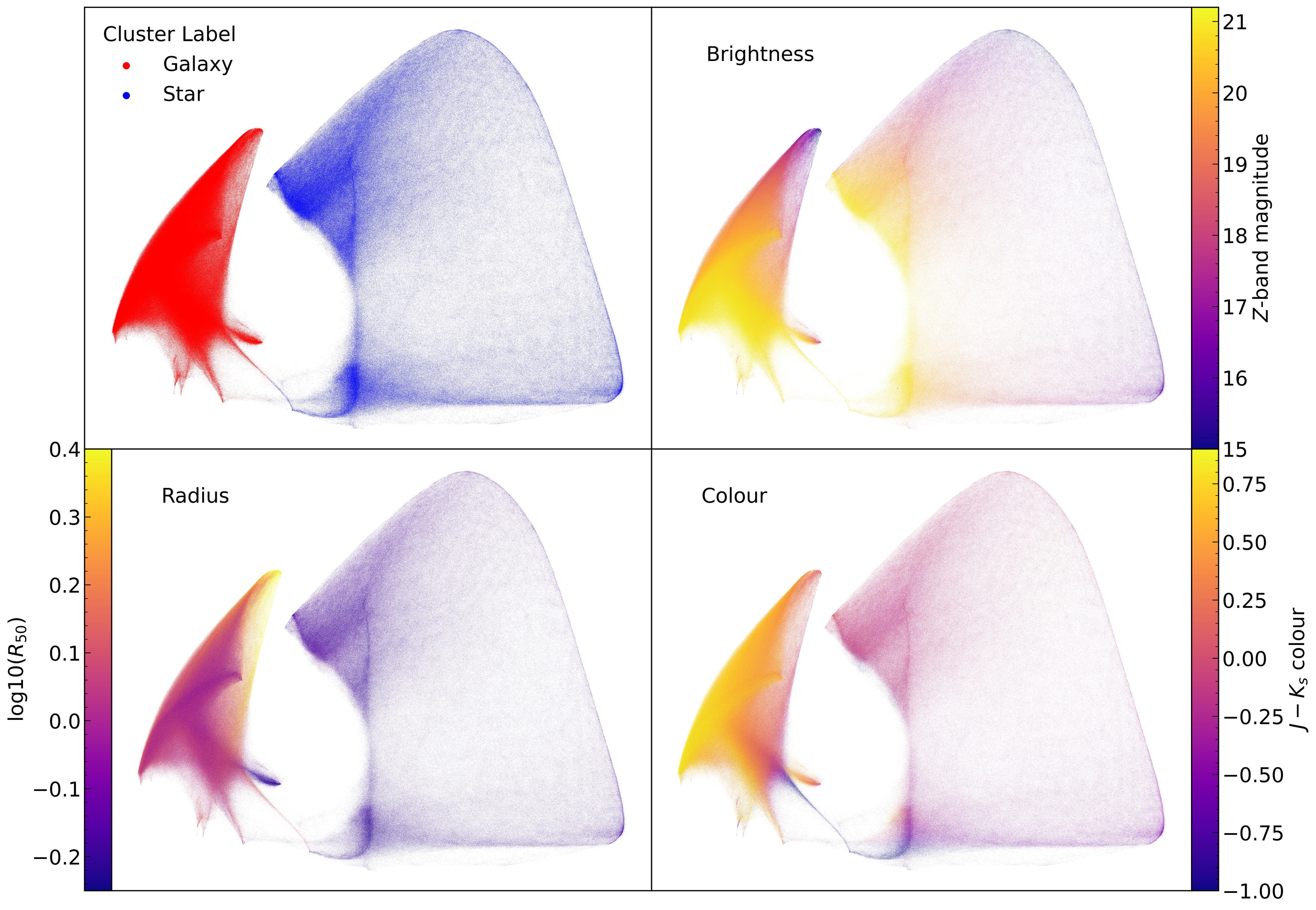}
    \caption{The same as Figure~\ref{fig:umap1_umap2_waves}, but coloured by different observed source properties. The top left shows the labels generated by \textsc{hdbscan}, clearly distinguishing the two clusters into stars and galaxies. The upper right plot is coloured by $Z$-band magnitude. The lower left plot is coloured by the log of the half-light radius. The lower right plot is coloured by the $J-K_s$ colour of each source. }
    \label{fig:umap1_umap2_all}
\end{figure*}

\begin{figure*}
    \centering
    \subfloat[\centering]{{\includegraphics[height=6.5cm]{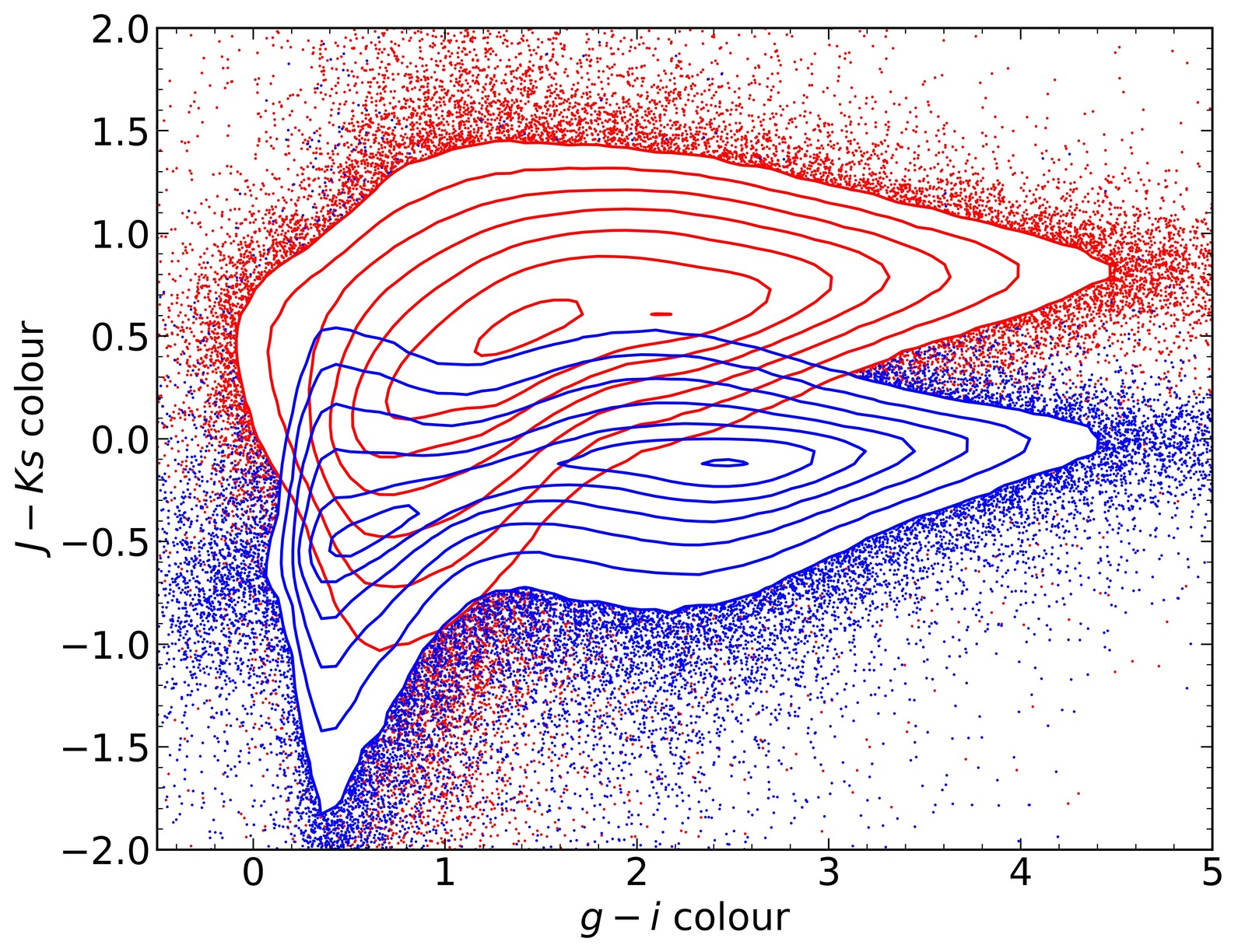} }}%
    \qquad
    \subfloat[\centering]{{\includegraphics[height=6.5cm]{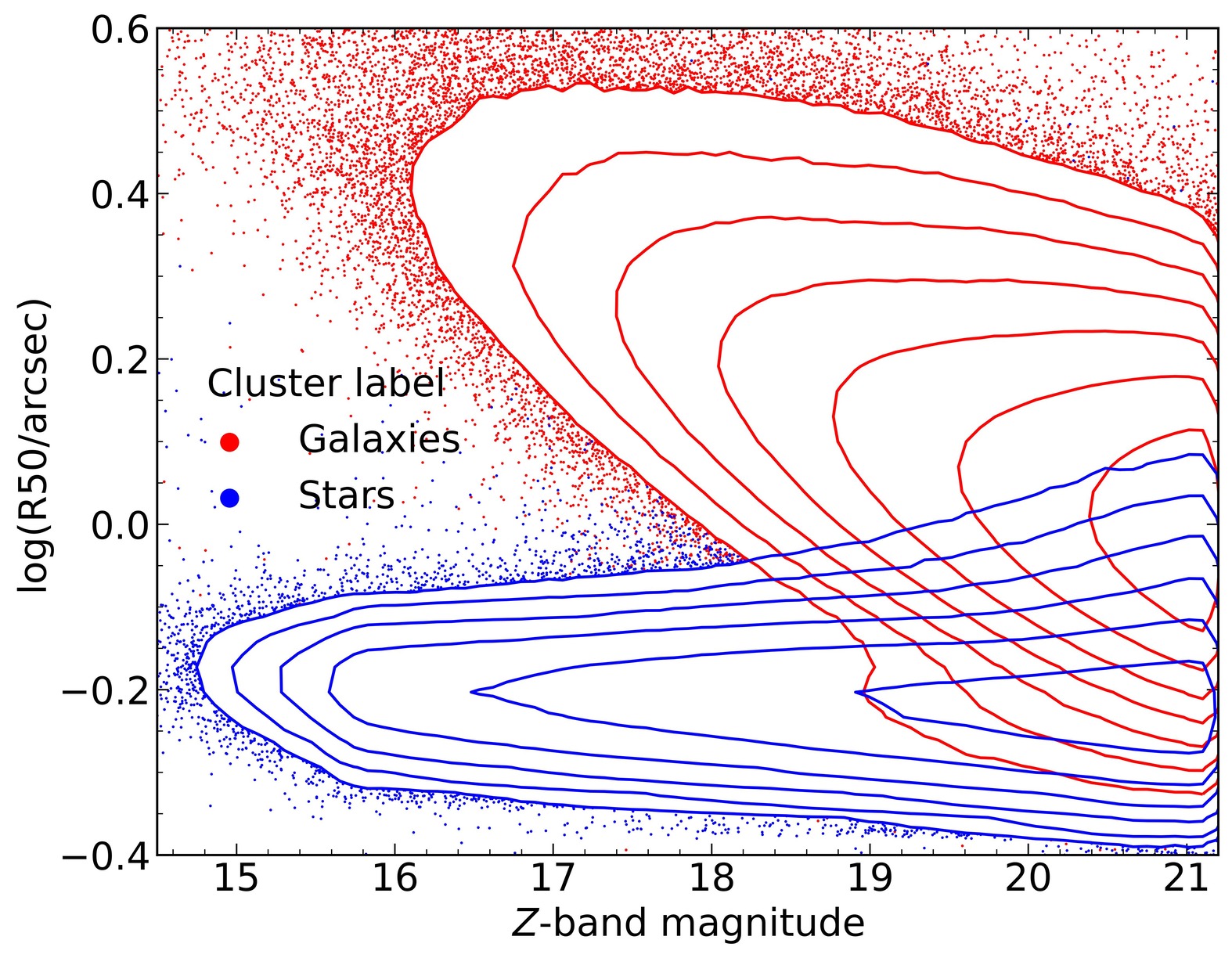} }}%
    \caption{Observable properties of the sources labelled stars and galaxies by our classifier. The left panel shows $J-K_s$ colour vs $g-i$ colour, and the right panel shows the log of half-light radius as a function of $Z-$band magnitude. Red and blue contours/points indicate galaxies and stars respectively. Contours are scaled logarithmically and the points show a random 10\% of each population.}
    \label{fig:cluter_properties}
\end{figure*}

Despite the UMAP embeddings being very distinctly clustered in 2 dimensions, we still employ an unsupervised cluster algorithm to precisely split the data. We use Hierarchical Density-Based Clustering (\textsc{hdbscan}; \citealp{mcinnes_hdbscan_2017}) for this purpose, a robust cluster-finding algorithm built on Density-Based Spatial Clustering of Applications with Noise, (\textsc{dbscan}: \citealp{ester_density-based_1996}). These algorithms work differently from other clustering algorithms such as $k$-means because they do not require a user-set number of clusters to work towards, and instead simply find areas of high density. $k$-means also require clusters to be `spherical' in feature space which is not always the case for features dervied from these surveys (\citealp{turner_reproducible_2019,holwerda_galaxy_2022}).

They work by first establishing a set of `key' points, those which lie in groups of at least $k$ points (set by the \texttt{min\_samples} argument in \textsc{hdbscan}), where points are linked if they lie within a set value, $\epsilon$. \textsc{hdbscan} improved upon \textsc{dbscan} by automatically finding an optimum $\epsilon$ instead of being user-set, making it more generalisable. \textsc{hdbscan} also uses a user-set \texttt{min\_cluster\_size} argument, a minimum threshold value below which clusters are discarded and the most sensitive parameter in this work.

\textsc{hdbscan} is used in astronomy for both extragalactic and galactic science. \cite{logan_unsupervised_2020} use \textsc{hdbscan} also for star-galaxy-QSO separation for SDSS using PCA and Random Forest as their feature selection. \cite{queiroz_starhorse_2023} use \textsc{hdbscan} in combination with t-SNE to group Gaia stars according to their spectra.

After several iterations of parameters, we settle on $\texttt{min\_samples}=$ \SI{1100}{} and $\texttt{min\_cluster\_size}=$ \SI{10000}{}, to reduce the chance of any fragmentation of the star and galaxy clusters. Any fragmented clusters are  manually merged into one of their parent clusters until two clusters remain. \textsc{hdbscan} also generates a probability for each source of belonging to its associated cluster.

We run \textsc{hdbscan} on the 10 dimensions of the data generated by UMAP, which results in each source being given a 0 or 1 label. We analyse the baseline classification for sources labelled 0 and 1, and change to `star' or `galaxy' accordingly. The resulting classification can be seen in the upper left panel of Figure~\ref{fig:umap1_umap2_all}. \textsc{hdbscan} has successfully distinguished the two main clusters, which we classify as stars and galaxies by comparing to the Baseline method in Figure~\ref{fig:umap1_umap2_waves}. Running \textsc{hdbscan} on the entire feature set is much more computationally expensive. We found that running \textsc{hdbscan} on a small patch of sky with the full feature set (without dimensionality reduction) is not as effective as applying UMAP first and then clustering.

For the sources with a $u$-band measurement, we find a small number of contradictory classifications running our method with and without the $u$-band. Of the \SI{9118285}{} sources classified as galaxies using $u$-band, \SI{9099596}{} of them (99.8\%) of them are also classified as galaxies \textit{without} using $u$-band. Similarly, of the \SI{4309681}{} sources classified as stars using $u$-band, \SI{4289080}{} of them (99.5\%) were also classified as stars \textit{without} using $u$-band. As stated before, for the sources which have contradictory labels, the classification using $u$-band takes priority.

We use \textsc{hdbscan}'s outlier detection (\texttt{outlier\_scores}) to give each source a likelihood that it is an outlier to the two main clusters generated by UMAP. \textsc{hdbscan} finds very few number of outliers in the UMAP feature space, with just 0.25\% of sources having an outlier score greater than 0.5. The majority of these are found in between the two UMAP clusters, connecting them.

\section{Classification performance using ground truth labels}
\label{classification_performance}

\begin{table}
    \centering
    \caption{The overall classifications of the WAVES-Wide sample made by our method and the baseline method.}
    \begin{tabular}{ r|c c c | c}
    \hline
    \hline
    \multicolumn{5}{c}{Total WAVES-Wide sample} \\
    Method & Galaxy & Star & Ambiguous & Total \\
    \hline
    \multirow{2}{*}{Cluster label} & 66.5\% & 33.5\% & & 100.0\% \\
     & \SI{9840496}{} & \SI{4961236}{} & & \SI{14802032}{} \\
    \multirow{2}{*}{Baseline} & 66.8\% & 22.2\% & 11.0\% & 100.0\% \\
     & \SI{9890177}{} & \SI{3288478}{} & \SI{1623377}{} & \SI{14802032}{} \\
     \hline
    \end{tabular}
    \label{table:overall}
\end{table}

The overall classifications of the \SI{14802032}{} sources is summarised in Table~\ref{table:overall}. If using the baseline classification scheme, we would select sources classified as galaxy or ambiguous in order to ensure the required completeness.  With UMAP/HDBSCAN (hereafter `cluster') classification, we will simply select targets classified as galaxies. Moving to this new classifier will result in \SI{1672758}{} fewer sources identified as galaxy or ambiguous, 11.3\% of the catalogue. We believe that this is due to the ambiguous class in the baseline method containing primarily stars. This is associated with a significant reduction in the size of the WAVES target catalogue. We are able to use the 4MOST exposure time calculator (4FS$\_$ETC) to estimate the reduction in fibre hours. 4FS$\_$ETC estimates exposure times based on the magnitude and assumed template for each source, as well as the required signal-to-noise ratio to recover a redshift measurement. We can compare the exposure times associated with the WAVES target catalogues using the baseline star-galaxy classifier and our new method to estimate the number of fibre hours saved. After the photometric redshift cuts have been made, we find that moving from the baseline to the cluster classification results in a reduction of \SI{69504}{} fibre hours for the WAVES wide region, and an additional \SI{69903}{} fibre hours reduction for the WAVES deep region.

We can tentatively gauge the fidelity of the classifier by looking at the properties of the stars and galaxies we have classified. Even though we are only looking at observed and not intrinsic properties of these sources, we can still use them for informative analysis. Figure~\ref{fig:cluter_properties} shows some observable properties of our classified stars and galaxies. The left panel shows a colour-colour plot, namely $J-K_s$ versus $g-i$: a plot classically used in colour-based star-galaxy separation (e.g. \citealp{ivezic_optical_2002,baldry_galaxy_2010}). The distinctive stellar locus can be seen, where stars have a roughly constant $J-K_s$ colour beyond $g-i>2$, and then reduces at smaller $g-i$. There is a significant amount of overlap at smaller $g-i$ colours. This indicates that a straight dividing line through $J-K_s=0.025$ to separate stars and galaxies as is done in the Baseline method would not be optimal. Similarly in the right panel of Figure~\ref{fig:cluter_properties}, we plot our sources' log of half-light radius as a function of $Z-$band magnitude. We see that in the bright regime, stars and galaxies can easily be classified according to their apparent radius. However, at the faintest magnitudes, the distinction becomes less clear. We see a slight uptick in apparent radius of stars at the faintest magnitudes, which could indicate some stars being misclassified as galaxies.

\subsection{Comparison to ground truth}

\begin{figure*}
    \centering
    \subfloat[\centering]{{\includegraphics[height=5.5cm]{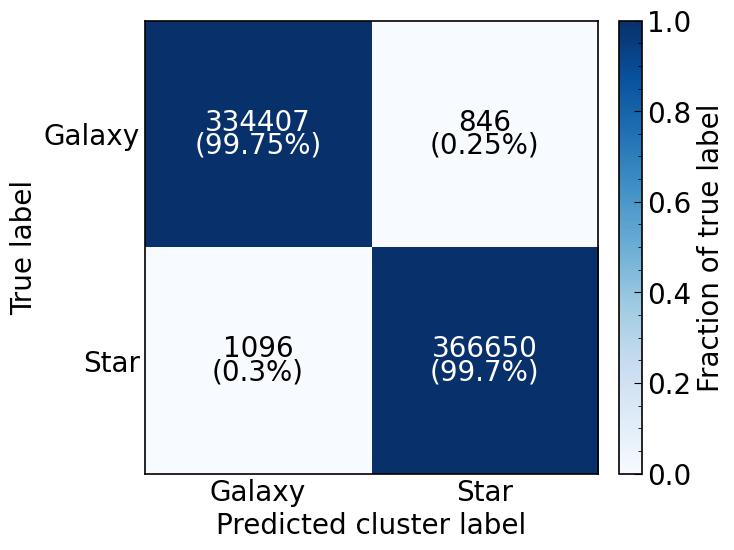} }}%
    \qquad
    \subfloat[\centering]{{\includegraphics[height=5.5cm]{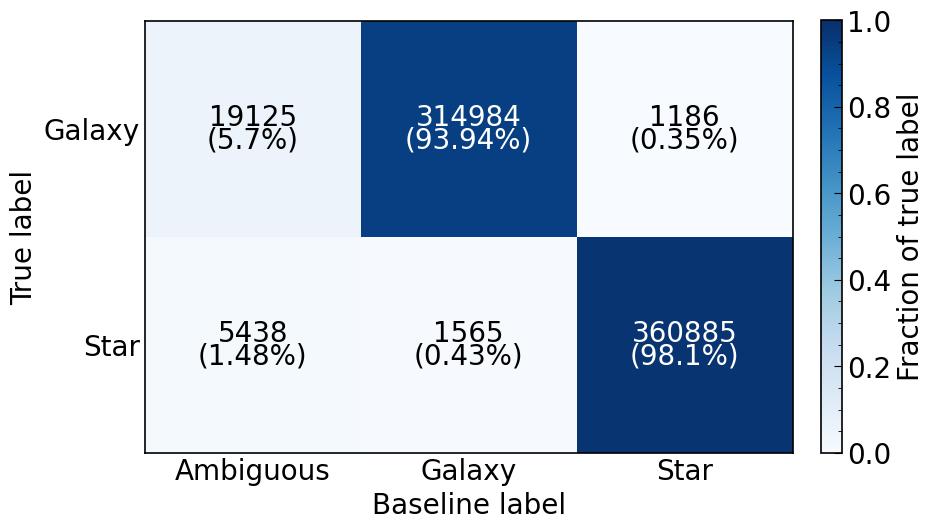} }}%
    \caption{The confusion matrices generated by the ground-truth dataset compiled in Section~\ref{groundtruth}. The left confusion matrix uses the labels generated by \textsc{hdbscan}, and the right confusion matrix uses the labels generated by the baseline algorithm described in Section~\ref{baselinemethod}. }
    \label{fig:confusion_matrix}
\end{figure*}

The most intuitive way to measure the fidelity of the new classification method is to compare with the ground truth classifications. Figure~\ref{fig:confusion_matrix} shows the confusion matrix between these labels. Of the \SI{335295}{} true galaxies, \SI{334407}{} of them have been correctly identified, an accuracy of 99.75\% compared to the baseline's accuracy of $93.94$\%. Similarly, of the \SI{367888}{} true stars, 99.70\% of them have been correctly identified, compared to the baseline accuracy of 98.10\%. Most of the losses of the baseline algorithm are due to the ambiguous class, which contains faint, harder to classify sources. Despite this, more galaxies are missed by the baseline algorithm, with \SI{1186}{} galaxies being misclassified as stars, compared to \SI{846}{} from our method. This implies that our method has been able to retrieve more galaxies (higher completeness) and includes fewer stars (higher purity) than the baseline method.

To quantify the fidelity of the method, we use the F1 score metric, which uses a harmonic mean of completeness and purity. In the context of identifying galaxies, we define a true positive (TP) as a true galaxy correctly identified, a false positive (FP) as a true star misclassified as a galaxy and a false negative (FN) as a true galaxy misclassified as a star. Purity is then defined as
\begin{equation}
    \mathrm{P = \frac{TP}{TP+FP}},
\end{equation}
the fraction of true positives to total positives.
Completeness is defined as
\begin{equation}
    \mathrm{C = \frac{TP}{TP+FN}},
\end{equation}
the fraction of positive prediction to the total number of positives in the sample. These are combined to form the F1 score
\begin{equation}
\label{eq:f1}
    \mathrm{F1 = 2 \ \frac{P \cdot C}{P+C}},
\end{equation}
the harmonic mean of purity and completeness.

In the context of an extragalactic target catalogue and taking true positives as correctly identified galaxies, our method achieves a purity of $0.9967 \pm 0.0021$, a completeness of $0.9975 \pm 0.0018$ and an F1 score of $0.9971 \pm 0.0018$. We derive the uncertainties of our metrics by taking the standard deviation of the metric across magnitude bins. The metrics' variability as a function of magnitude is explored later.

Using the baseline algorithm, it can be assumed that all ambiguous objects would be part of the target catalogue for an extragalactic survey. This is done to maximise the completeness, a metric critical for the construction of group catalogues (\citealp{robotham_galaxy_2011, tempel_flux-_2014,tully_galaxy_2015}). Assuming this, the baseline algorithm achieves a purity of $0.9795 \pm 0.0172$, a completeness of $0.9965 \pm 0.0026$ and an F1 score of $0.9879 \pm 0.0088$.

\begin{table}
    \centering
    \caption{Overall purity, completeness and F1 scores of the two methods, using all of the ground-truth labels.}
    \begin{tabular}{ r|c c c }
    \hline
    \hline
    Method & Purity & Completeness & F1 \\
    \hline
    Cluster label & $0.9967 \pm 0.0021$ & $0.9975 \pm 0.0018$ & $0.9971 \pm 0.0018$ \\
    Baseline\tablefootnote{Taking all ambiguous sources to be galaxies.} & $0.9795 \pm 0.0172$ & $0.9965 \pm 0.0026$ & $0.9879 \pm 0.0088$ \\
    \hline
    \end{tabular}
\end{table}

We observe that whilst the completeness stays the same between the two methods (i.e. roughly the same number of galaxies are retrieved), the purity increases from $0.9795 \pm 0.0172$ to $0.9967 \pm 0.0021$ and consequently the F1 score increases too from $0.9879 \pm 0.0088$ to $0.9971 \pm 0.0018$. In the case of a spectroscopic extragalactic survey, this would imply fewer stars in the targeting catalogue and therefore less wasted fibre hours.

In addition, we find that of the $z<0.2$ galaxies in the ground truth sample (the photo-z limit of WAVES-Wide), we identify 99.75\% of them as galaxies, compared to 88.50\% of them being classified as galaxies by the baseline algorithm. This highlights a major improvement in the star-galaxy separation in the context of WAVES-Wide. For galaxies with redshifts $z<0.8$ (the photo-z limit of WAVES-Deep), we correctly identify 99.82\%, compared to 94.85\% from the baseline algorithm.

\begin{table}

\centering
\caption{The F1 scores achieved by a selection of pre-processing and clustering methods, run on a one tenth representative sample of WAVES-Wide.}
    \begin{tabular}{ c | c c c}
      \hline
      \hline
      \diagbox[width=\dimexpr \textwidth/8+2\tabcolsep\relax, height=1cm]{ Clustering }{Preprocessing}
                   & None & PCA & UMAP \\
        \hline & \\[-1.5ex]
        $k-$means & $0.621 \pm 0.194$ & $0.623 \pm 0.193$ & $0.997 \pm 0.004$ \\
        & & & \\
        \textsc{hdbscan} & $0.935 \pm 0.046$ & $0.947 \pm 0.037$ & $0.997 \pm 0.004$ \\[1.5ex]
        \hline
    \end{tabular}
\label{table:process_cluster}
\end{table}

To demonstrate the importance of UMAP in our method, we attempt star-galaxy classification with a number of preprocessing and clustering methods. We create a random subset of WAVES-Wide which is a 10th of the size but fully representative of the entire catalogue in all dimensions. For our preprocessing, we use UMAP, PCA and also attempt clustering without any preprocessing. We use $k-$means clustering and \textsc{hdbscan} as our clustering methods.

The results can be seen in Table~\ref{table:process_cluster}, in which we show the F1 scores achieved from the ground truth sample by combining the various preprocessing and clustering methods. We find that when using UMAP as our preprocessing method, both $k-$means and \textsc{hdbscan} both achieve the same F1 score, as UMAP makes the clustering trivial (shown in Figure~\ref{fig:umap1_umap2_waves}). They achieve F1 scores of $0.997\pm0.004$, compared to the F1 scores of $0.947\pm0.037$ and $0.935\pm0.046$ achieved by \textsc{hdbscan} when using no preprocessing and PCA respectively. This indicates that is indeed UMAP that is driving the performance of our star-galaxy classification, and not \textsc{hdbscan}.

\subsection{As a function of magnitude}
\begin{figure}
    \includegraphics[width=\columnwidth]{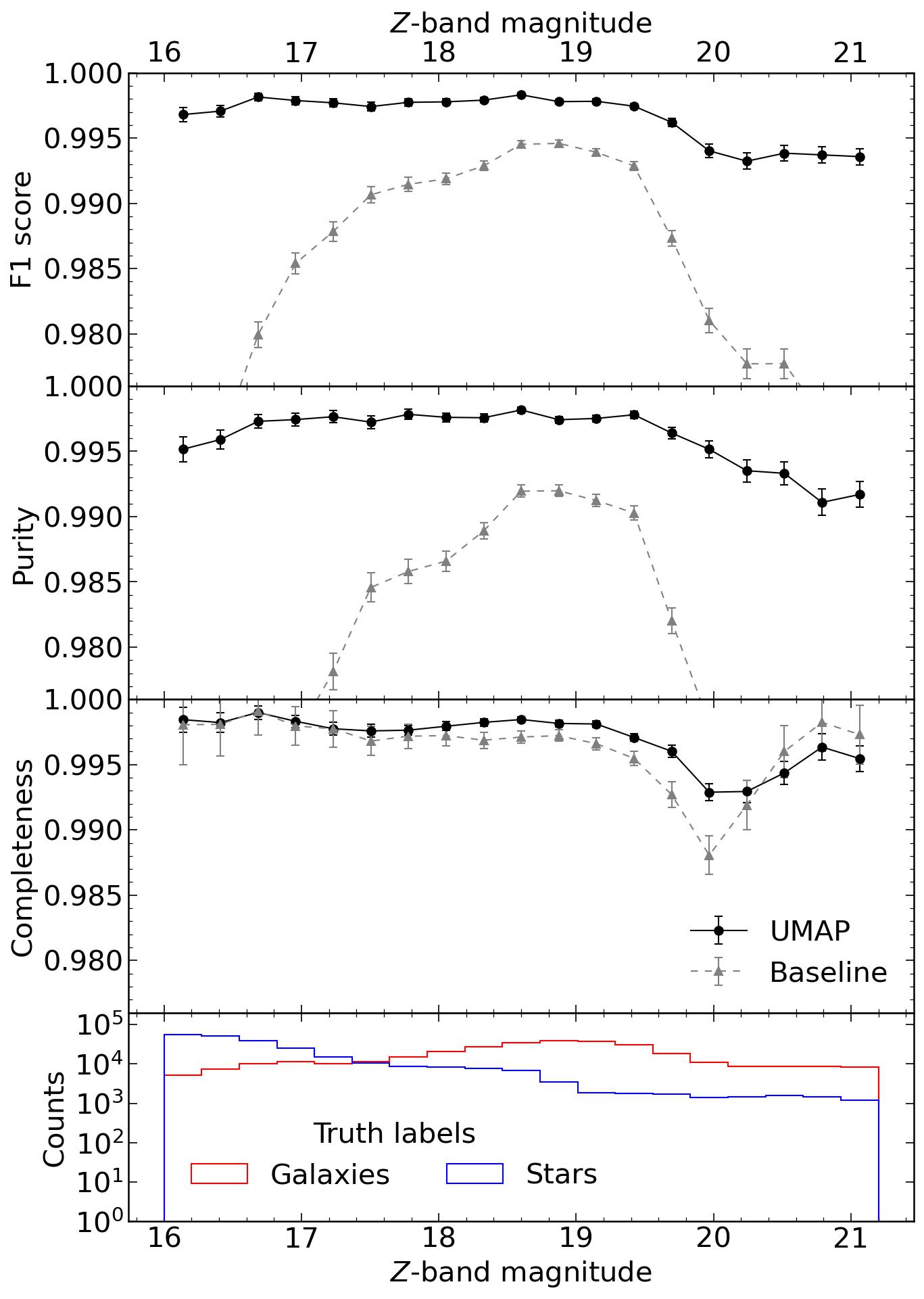}
    \caption{The F1 score (Equation~\ref{eq:f1}), purity and completeness for identifying galaxies as a function of $Z$-band magnitude. The solid line shows the performance of our algorithm, compared to the baseline algorithm in grey. Errors are based on Poisson statistics from the number of ground truth sources we have, shown in the bottom plot.}
    \label{fig:z_mag_f1}
\end{figure}

We can also measure the fidelity of our method as a function of different source properties. For example, Figure~\ref{fig:z_mag_f1} shows how the F1 score (Equation~\ref{eq:f1}), purity and completeness for classified galaxies varies as a function of $Z$-band magnitude compared to the baseline algorithm. Across all magnitudes, the purity produced by our method exceeds the baseline algorithm, suggesting our sample is far less contaminated, especially at very faint ($Z>20$) and very bright ($Z<17$) magnitudes. The purity from our method never dips below 0.99, whereas the purity from the baseline method is 0.970 at the faintest magnitudes. The completeness of both our method and the baseline algorithm hover around 0.995, with a dip at about $Z \sim 20$, which we believe is caused by a shift from the GAMA to DESI regime in the ground truth sample. This results in the total F1 score never reaching below 0.990, and consistently above the baseline method at all magnitudes.

\subsection{Classification probabilities}
\begin{figure}
    \includegraphics[width=\columnwidth]{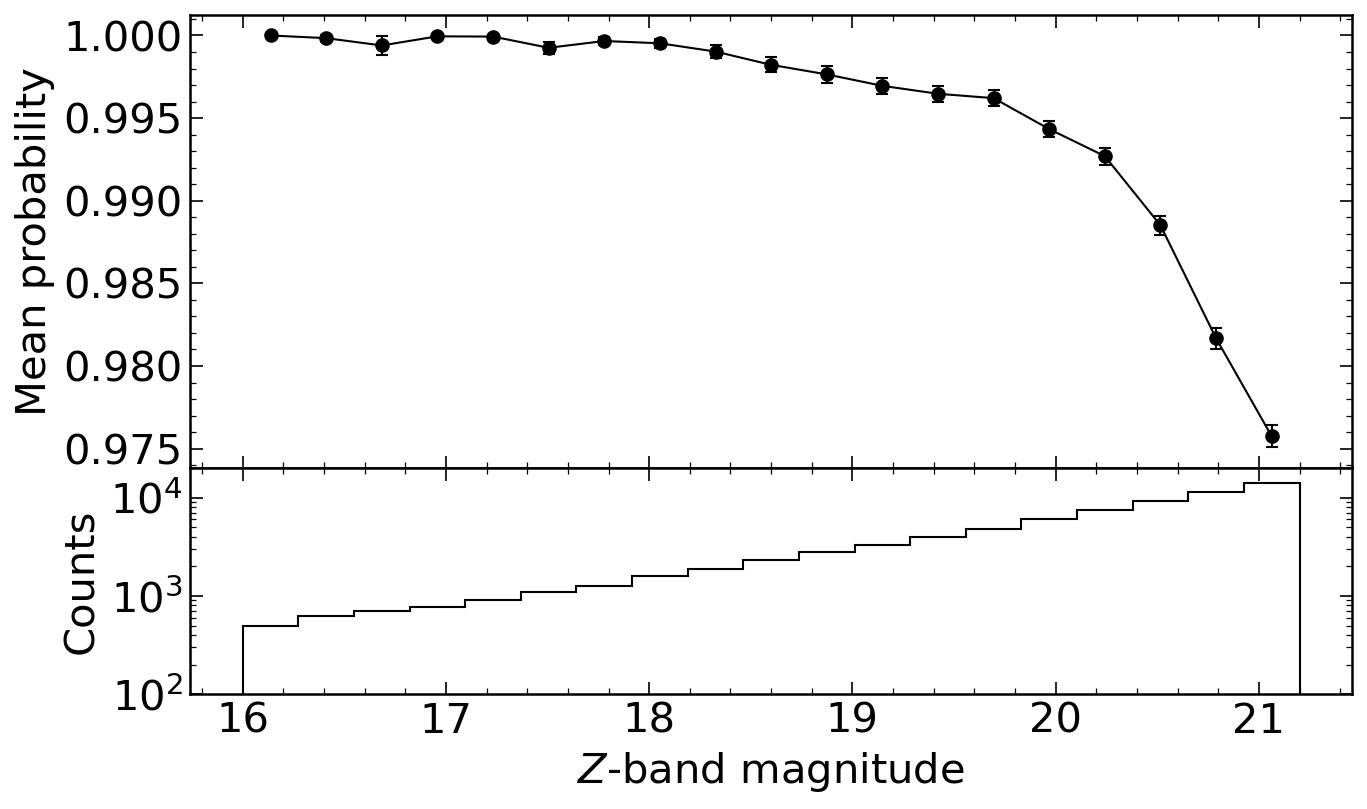}
    \caption{The mean probability of classifications in bins of $Z$-band magnitude, calculated by perturbing the flux values of sources by their uncertainties and measuring the number of contradictory classifications. The bottom panel shows the number of sources per magnitude bin. In this analysis, we use a representative sample of WAVES-Wide, one hundredth of the size.}
    \label{fig:probability_magnitude}
\end{figure}

We also investigate how the uncertainties in the flux measurements of each source can change the classification. The flux errors are calculated in \textsc{ProFound} from sky subtraction and sky rms, with errors being higher in areas of great noise, e.g. near bright objects. We create a representative sample of WAVES-Wide, one hundredth the size, for ease of computation.
We assume Gaussian uncertainties, and perturb the fluxes of each source by their flux uncertainty 100 times, and classify the sources as stars and galaxies as before, keeping track of the classification of each source. We then assign each source a classification probability, based on the number of contradictory classifications after the 100 perturbations.

The results can be visualised in Figure~\ref{fig:probability_magnitude}, in which we show the mean of the classification probabilities as a function of $Z$-band magnitude. As expected, the probabilities decrease towards fainter magnitudes, with an average probability of $0.9757\pm0.0007$ in the faintest magnitude bin. This implies that on average, 2.4 out of every 100 perturbations will results in a contradictory classification at $Z=21.2$. The mean probability of the overall sample is $0.9898 \pm 0.0002$. When comparing to our ground truth catalogue, we find we find a standard deviation of 0.0006 in F1 score based on the 100 perturbations. This however is likely to be an underestimate due to the ground truth catalogue being biased towards bright objects. With enough computing power, we could use this method for the entire WAVES-Wide sample to give every source a classification probability. This would give another parameter for the 4MOST target catalogue, with fibre allocation being prioritised for galaxies with high probabilities.

\section{Classification of challenging galaxies}
\label{challening_galaxies}

In addition to assessing the ground truth sample as a whole, we also look at galaxies that could be challenging for star-galaxy classification. These include quasars, compact galaxies and low-surface brightness galaxies. We do this to ensure that the completeness of the star-galaxy classifier is effective even with these fringe cases. Figure~\ref{fig:exemplar} shows nine of these challenging galaxies using KiDS and VIKING photometry. All have been correctly classified as galaxies by our algorithm, but some are misclassified by the baseline algorithm.

\subsection{Quasars}
\label{quasars}

Firstly, we assess how well our star-galaxy separation performs at classifying quasars. Quasars or QSOs are galaxies with an extremely luminous active galactic nucleus, meaning they can be observed even at high redshifts. Like stars, they appear almost as point sources, meaning their classification from stars is a difficult task, and careful analysis is required.  Quasars can be identified using infrared and optical photometry (\citealp{assef_wise_2018,chaussidon_target_2023}), UV excess (\citealp{richards_spectroscopic_2002}), photometric variability (\citealp{macleod_quasar_2011}) and X-ray excess (\citealp{maccacaro_x-ray_1984}).

For our true quasar sample, we use those identified spectroscopically in the DESI EDR. Quasars are identified in DESI through their spectral classification pipeline using Redrock template fitting software (\citealp{guy_spectroscopic_2023}). Redrock fits a set of PCA templates to each source at every redshift, and the classification and redshift is returned with the smallest $\chi^2$ fit, regardless of the target selection type. Many galaxies with AGN are classified as quasars in the EDR (\citealp{desi_collaboration_early_2023}), producing a `bump' of sources at $z<0.5$. We filter these out by selecting only sources with a PSF morphological type, increasing the median redshift from $z=1.42$ to $z=1.61$.

We are able to cross-match \SI{5929}{} quasars in the North region. The results of the two classifiers can be seen in Table \ref{qso_table}. Of the \SI{5929}{} quasars, our method was able to correctly identify \SI{5494}{} or 92.7\% of them as galaxies. This is compared to the baseline algorithm, which was only able to identify \SI{1418}{} or 23.9\% of them as galaxies, with the majority being classified as ambiguous sources. Even if all ambiguous sources were observed (an additional 1.2 million sources), our method would have still identified more quasars in these data.

Further study could be conducted in using \textsc{hdbscan} to create an additional quasar cluster\footnote{``Clusters'' in the sense of the classification algorithm} separated from the galaxy and star clusters. However, for the purpose of this work, we are only interested in separating them from stars. We mention earlier that any quasars correctly identified as galaxies should be filtered out of the WAVES-Wide target catalogue due to the $z<0.2$ photo-z selection. We find that of the DESI EDR quasars with spectroscopic redshifts of $z>0.2$, $\sim 95\%$ of them are filtered out using the photo-z selection.

\begin{table}
    \centering
    \caption{The classification of DESI EDR quasars achieved by the method described in this paper and the baseline method.}
    \label{qso_table}
    \begin{tabular}{ r|c c c | c}
    \hline
    \hline
    \multicolumn{5}{c}{DESI EDR quasars} \\
    Method & Galaxy & Star & Ambiguous & Total \\
    \hline
    \multirow{2}{*}{Cluster label} & 92.7\% & 7.3\% & & 100.0\% \\
     & \SI{5494}{} & 431 & & \SI{5929}{} \\
    \multirow{2}{*}{Baseline} & 23.9\% & 12.3\% & 63.8\% & 100.0\% \\
     & \SI{1418}{} & 729 & \SI{3782}{} & \SI{5929}{} \\
     \hline
    \end{tabular}
\end{table}

\subsection{Compact galaxies}
\label{compact}

Next, we measure the effectiveness of our star-galaxy separation in identifying compact galaxies. Compact galaxies are characterised as having a high concentration of their mass and luminosity in a small, central area of the galaxy. Their compact profile makes them difficult to distinguish from stars using only morphological data, so colour information is necessary.

We utilise the metric  $\Sigma_{1.5} = \mathrm{log}(M_{*} / \mathrm{M}_{\odot}) - 1.5\mathrm{log}(r_{50} / \mathrm{kpc}) $ to quantify compactness, first outlined in \cite{barro_candels_2013} for the CANDELS survey and later in \cite{baldry_compact_2021} for SDSS, where $M_{*}$ is stellar mass and $r_{50}$ is the physical half-light radius of the galaxy. The metric essentially measures the deviation from the size-mass relation of high-mass, quiescent galaxies, and is measured in units of $\mathrm{M_\odot kpc^{-1.5}}$. A greater $\Sigma_{1.5}$ value indicates a more compact galaxy.

We use stellar masses from GAMA DR4 (\citealp{driver_galaxy_2022}), which uses code first outlined in \cite{taylor_galaxy_2011}, but has been developed into DR4 using Source Extractor photometry from \cite{driver_galaxy_2016}, matched aperture photometry from \textsc{lambdar} (\citealp{wright_galaxy_2016}) and \textsc{ProFound} photometry from \cite{bellstedt_galaxy_2020}. It also uses GAMA spectroscopic redshifts combined with the apparent half-light radius to calculate the physical half-light radius. We limit the GAMA sample to redshifts less than 0.6 to avoid quasars, and define our compact sample as galaxies with the highest 0.5\% of $\Sigma_{1.5}$ values. This leaves us with a sample of 878 galaxies with a mean compactness value of $\Sigma_{1.5} =  \SI{10.28}{\mathrm{M_\odot kpc^{-1.5}}} $, compared to $\Sigma_{1.5} = \SI{9.34}{\mathrm{M_\odot kpc^{-1.5}}}$ for the whole GAMA sample.

Table \ref{compact_table} shows the performance of our star-galaxy separation and the baseline algorithm on the compact GAMA sample. We correctly identify 84.6\% of the compact galaxy sample as galaxies, and misclassify 15.4\% of them as stars.  This is compared to the baseline algorithm classifying 81.8\%, 21.2\% and 6.2\% of the compact galaxies as galaxies, stars and ambiguous sources respectively. This is a slight improvement in the retrieval of compact galaxies, unless all ambiguous sources are also considered. We do make the caveat that GAMA is magnitude limited to $r<19.8$, and its input catalogue likely omitted more compact galaxies in its own star-galaxy separation (\citealp{baldry_galaxy_2010}). We also note that if all ambiguous objects were observed, this would result in more compact galaxies being retrieved (93.2\% compared to 84.5\%), although this increase in completeness comes at the cost of observing an additional 1.2 million sources.

\begin{table}
    \centering
    \caption{The classification of compact galaxies described in Section~\ref{compact} achieved by the method described in this paper and the baseline method.}
    \label{compact_table}
    \begin{tabular}{ r|c c c | c}
    \hline
    \hline
    \multicolumn{5}{c}{Compact GAMA galaxies} \\
    Method & Galaxy & Star & Ambiguous & Total \\
    \hline
    \multirow{2}{*}{Cluster label} & 84.6\% & 15.4\% & & 100.0\% \\
     & 743 & 135 & & 878 \\
    \multirow{2}{*}{Baseline} & 81.8\% & 12.1\% & 6.2\% & 100.0\% \\
     & 718 & 106 & 54 & 878 \\
     \hline
    \end{tabular}
\end{table}

\subsection{Low surface brightness galaxies}
\label{lsb}

Finally, we look at low surface brightness galaxies. These are galaxies with a faint overall brightness over their area, making them difficult to detect in photometric data. They span a wide range of colours (\citealp{greene_nature_2022}), with their bluer members actively forming stars. There is a concern that some of these low surface brightness galaxies have extreme colours, and thus could be missed by simple colour-based star-galaxy separation methods. They may also have unreliable photometric redshifts which one may also use in a method for star-galaxy separation.

We calculate the effective surface brightness $\mu_{z,50}$ for each DESI and GAMA galaxy using their area calculated from their half-light radius $R_{50}$ and $Z$-band magnitude. We limit the sample to have a redshift of $z<0.2$ and define the low surface brightness sample as having the smallest 0.5\% $\mu_{z,50}$ values. This sample comprises 444 galaxies, with a mean $\mu_{z,50}$ surface brightness value of \SI{22.93}{mag \ arcsec^{-2}}, compared to the mean surface brightness of \SI{20.28}{mag \ arcsec^{-2}} of the rest of the $z<0.2$ GAMA and DESI samples.

The results can be seen in Table \ref{lsb_table}. Of the 444 low surface brightness galaxies in the sample, we correctly identify 442 of them as galaxies. This is compared to the baseline algorithm which classifies just 261 as galaxies and the rest as ambiguous sources. Their large apparent radii mean they lie in the galaxy region in the lower plot of Figure~\ref{fig:r_j-k_r50}, but they exhibit a wide range of colours. The 25th and 75th percentile of their $J-K_s$ colours are -0.31 and 0.28 respectively, compared to that of the ground truth galaxy sample of 0.12 and 0.32. 50.2\% of the low-surface brightness galaxies have a $J-K_s$ colour less than the Baseline cutoff of 0.025, placing them in the star region of the upper panel (colours versus magnitude) of Figure~\ref{fig:r_j-k_r50}, thus classing them as ambiguous despite their large apparent radii.

\begin{table}
    \centering
    \caption{The classification of low surface brightness galaxies described in Section~\ref{lsb} achieved by the method described in this paper and the baseline method.}
    \label{lsb_table}
    \begin{tabular}{ r|c c c | c}
    \hline
    \hline
    \multicolumn{5}{c}{Low surface brightness galaxies} \\
    Method & Galaxy & Star & Ambiguous & Total \\
    \hline
    \multirow{2}{*}{Cluster label} & 99.5\% & 0.5\% & & 100.0\% \\
     & 442 & 2 & & 444 \\
    \multirow{2}{*}{Baseline} & 58.8\% & 0.0\% & 41.2\% & 100.0\% \\
     & 261 & 0 & 183 & 444 \\
     \hline
    \end{tabular}
\end{table}

\section{Classification performance using angular two-point correlation function}
\label{clustering}

This section utilises the angular two-point correlation function, $\omega(\theta)$, for assessing star-galaxy separation in the population as a whole. This tool can be useful in differentiating stars from galaxies, given their distinct clustering patterns. Galaxies typically cluster following a power-law on small angular scales,
\begin{equation}\label{eq:wt_power}
    \omega(\theta) = A_{\omega} \theta^{\delta},
\end{equation}
with $\delta \approx -0.8$ \citep{Groth1977,Coil2012}.
In contrast to galaxies, stars exhibit a different clustering pattern, as evidenced by their angular two-point correlation function. Stars typically do not demonstrate the large-scale clustering patterns that are observable in galaxies.  Unlike galaxies, stars are essentially randomly distributed apart from a large-scale gradient towards the Galactic centre and plane. This difference in clustering pattern allows an assessment of star/galaxy classification, independent of the ground-truth dataset.


\subsection{Methodology}
In this test, we use the Landy \& Szalay estimator (LS; \citealp{landy_bias_1993}), which is the most commonly used two-point correlation function estimator \citep{Coil2012}. It is defined as
\begin{equation}\label{eq:LS_est}
    \omega(\theta)=\frac{1}{RR} \left[
    DD \left( \frac{n_R}{n_D}\right)^2
    - 2DR \left( \frac{n_R}{n_D} \right)
    + RR
    \right],
\end{equation}
where $DD$ and $DR$ are the pair count of galaxies in each separation bin in the data catalogue and between the data and random catalogues, and $RR$ is the pair count for the random catalogue. $n_D$ and $n_R$ are the number densities of galaxies in the data and random catalogues.
In \citet{Kerscher2000}, it has been shown that the LS estimator handles edge corrections better than other estimators on large scales and matches the performance on small scales; also the size of the random catalogue affects it less than other estimators \citep{Coil2012}.

A "random catalogue" was constructed to mirror the angular coverage of our observational data, maintaining the same sky coverage but populated with randomly distributed points. We used Pymangle, a Python implementation derived from the Mangle C++ package \citep{swanson_methods_2008}, to generate and distribute these random points and apply the star mask that was created based on Gaia stars, where the mask radius is a function of the magnitude of each star \citep{bellstedt_galaxy_2020} from Gaia DR2 \citep{Gaia_collaboration_Gaia_2018}. This approach led to the production of random samples that contained ten times more points than our original catalogue, ensuring a comprehensive and statistically sound basis for our angular analysis. In this research, we used the TreeCorr Python package \citep{jarvis_skewness_2004} to calculate two point correlation functions.

Our test involves estimating the angular correlation function for objects classified by our method and the baseline method into five categories for star-galaxy separation:
\begin{enumerate}
    \item Objects identified as galaxies by both our method and the baseline method.
    \item All objects identified as galaxies by our method.
    \item Objects identified as galaxies by our method but ambiguous by the baseline method.
    \item Objects identified as stars by both methods.
    \item Objects identified as stars by our method but ambiguous by the baseline method.
\end{enumerate}

\subsection{Results}

\begin{figure*}
    \includegraphics[width=\textwidth]{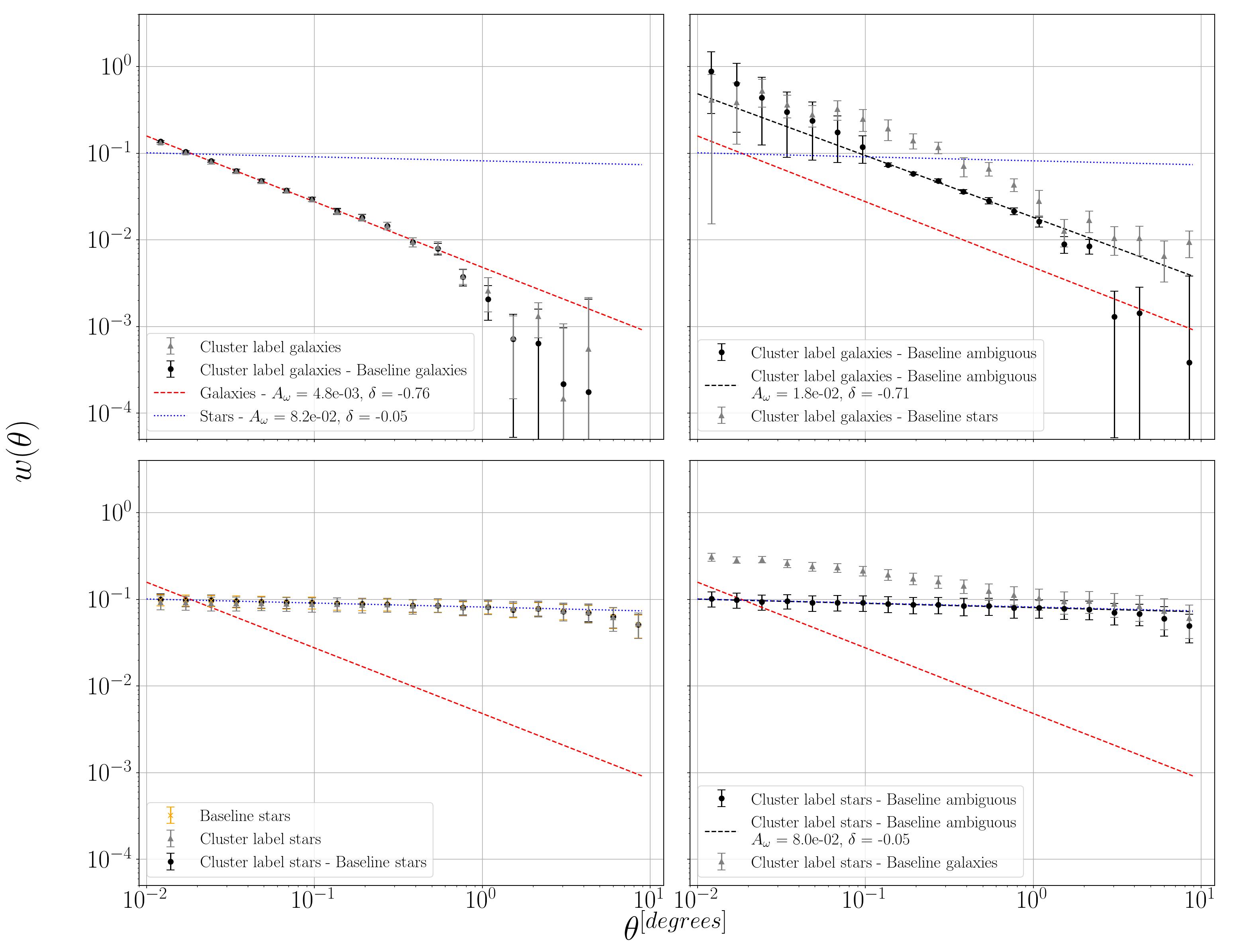}
    \caption{The angular two-point correlation function, $\omega(\theta)$, for various object classifications: galaxies, stars categorised by our method and the baseline method, and those objects which baseline method classified them as ambiguous. The red and blue dashed lines correspond to the best-fit power-law models for galaxies and stars for which both methods agree in their classifications.}
    \label{fig:wtheta}
\end{figure*}

\begin{table}
    \centering
    \caption{Power law fit (equation \ref{eq:wt_power}) for angular separation $\theta<0.9^{\circ}$.}
    \label{table:wtheta_fits}
    \begin{tabular}{c c c | c c}
    \hline
    \hline
    \multicolumn{5}{c}{Power law fit parameters} \\

    & \multicolumn{2}{c|}{Classification} & \multirow{2}{*}{$A_\omega$}   & \multirow{2}{*}{$\delta$} \\
    & Cluster   & Baseline      &                       &                        \\

    \hline

    i & Galaxy & Galaxy & $\left(4.8\pm{0.2}\right)\times 10^{-3}$ & $-0.76\pm{0.01}$ \\

    ii & Galaxy & All & $\left(4.7\pm{0.3}\right)\times 10^{-3}$ & $-0.76\pm{0.02}$ \\

    iii & Galaxy & Ambiguous & $\left(1.8\pm{0.1}\right)\times 10^{-2}$ & $-0.71\pm{0.02}$ \\

    \hline

    iv & Star & Star & $\left(8.2\pm{0.1}\right)\times 10^{-2}$ & $-0.05\pm{0.02}$ \\

    v & Star & Ambiguous & $\left(8.0\pm{0.1}\right)\times 10^{-3}$ & $-0.05\pm{0.03}$ \\

    \hline
    \end{tabular}
\end{table}

The angular correlation functions for the above sub-samples are plotted in Figure~\ref{fig:wtheta} and the parameters of power-law fits to scales $\theta < 0.9^\circ$ are given in Table~\ref{table:wtheta_fits}.
The differences in clustering pattern for galaxies and stars for which both classification methods agree are clear.
For sources classified as galaxies by both methods, for angular separation less than $0.9^{\circ}$, the fit parameters, a clustering amplitude \( A_{\omega} \) of \( (4.8 \pm 0.3) \times 10^{-3} \) and a slope \( \delta \) of \( -0.76 \pm 0.01 \), suggest a robust clustering measurement, in line with the established large-scale structure of galaxies. Conversely, sources classified as stars by both methods exhibit a markedly shallower correlation function, with an amplitude of \( (8.1 \pm 0.1) \times 10^{-2} \) and a slope \( \delta \) of \( -0.05 \pm 0.01 \). The red and blue dashed lines in Figure~\ref{fig:wtheta} represent the power-law fits for galaxies and stars, respectively, on which both methods agree in their classifications.

For sources identified as galaxies by our method, shown as grey points in the top left plot of Figure~\ref{fig:wtheta}, the clustering is similar to those sources agreed upon by both methods. This suggests that our classification of sources as galaxies remains uniform and consistent, irrespective of how the baseline method classifies them.

For objects identified as stars by our method and ambiguous by the baseline method, the bottom right plot in Figure~\ref{fig:wtheta} reveals that these sources cluster similarly to stars. This suggests that our method effectively recognises star-like properties in these objects, which the baseline method categorises as ambiguous, underscoring our method's effectiveness in definitively identifying stars.

For objects identified as galaxies by our method, yet deemed ambiguous by the baseline method, the top right plot in Figure~\ref{fig:wtheta} presents the angular correlation function. This function exhibits a slope similar to that of galaxies, and it is parallel to the power-law fit, but there is an observed offset from the fit for sources classified as galaxies by both methods. This deviation is explored in Appendix \ref{tiling_section}, where it is shown to arise from systematic variations in the seeing correction in the WAVES input catalogue adversely affecting the baseline classification.

Based on the results of this analysis, we can conclude that our method effectively identifies ambiguous sources that the baseline method was unable to classify definitively. This demonstrates the robustness and precision of our approach in handling cases where the baseline classification was uncertain.

\section{Summary}

We use unsupervised machine learning for star-galaxy separation of the WAVES-Wide input catalogue, and compare our results against a baseline method using a number of verification methods.

\begin{itemize}

\item[i)] We construct a catalogue of ground truth data for verification, formed from Gaia stars with high signal-to-noise parallax measurements, and stars and galaxies from GAMA, SDSS and DESI EDR spectroscopy. This gives us a sample of truth data even at faint magnitudes down to $Z<21.2$. Figure~\ref{fig:z_mag_count} demonstrates that the quantity of ground truth data at these faint magnitudes is orders of magnitudes smaller than the number of sources we wish to classify, giving us motivation to deviate from supervised machine learning in fear of bias from a training set.

\

\item[ii)] We utilise photometric features derived from the source-finding software \textsc{ProFound}, including KiDS and VIKING magnitudes, their colours, and sources' radii and axial ratios. A feature space is formed and reduced using UMAP, a non-linear dimensionality reduction algorithm, and then clustered into stars and galaxies using \textsc{hdbscan}.

\

\item[iii)] Our method classifies \SI{1672758}{} fewer sources as galaxy or ambiguous compared to the baseline method, or 11.3\%, which is a associated with an approximate reduction of ~\SI{70000}{} fewer 4MOST fibre hours after photometric redshift cuts, and fewer suprious stars. We compare the classification of the ground truth data to a baseline method, used in the star-galaxy separation for GAMA DR4 (\citealp{bellstedt_galaxy_2020}), which uses a combination of morphological and colour classification. The full results can be seen in the confusion matrices of Figure~\ref{fig:confusion_matrix}. They show a 5.86\% increase in the number of galaxies being correctly recovered, and a smaller amount being falsely classified as star or ambiguous sources. This can be quantified by an increase in F1 score from $0.9879 \pm 0.0088$ to $0.9971 \pm 0.0018$. We also demonstrate that the F1 score is better for our method than the baseline method across all magnitudes in Figure~\ref{fig:z_mag_f1}. This is mainly due to a major improvement in the purity of the sample.

\

\item[iv)] We assess the effectiveness of our method with `challenging' galaxies, including quasars, compact galaxies and low surface brightness galaxies. We find that our method outperforms the baseline method for all three types. The baseline method only manages to classify 36.1\% of quasars as galaxies, with most being classified as ambiguous due to them being point sources. This compares to our method identifying 95.1\% of them correctly as galaxies. There is a minor increase in effectiveness in identifying compact galaxies, and a significant improvement in identifying low surface brightness galaxies due to their extraordinary colours.

\

\item[v)] Finally, we investigate the angular clustering of our sources and baseline sources. We find that the angular clustering of our stars and galaxies are consistent with what we expect. We also find that even for sources classified as ambiguous by the baseline method, if our method classifies these sources as stars or galaxies, then they exhibit the expected clustering.

\end{itemize}

Our method of unsupervised star-galaxy separation for the WAVES-Wide target catalogue shows promising results, improving on the baseline method in both purity and completeness, and saving valuable fibre hours on the 4MOST instrument. We hope the use of UMAP and other unsupervised machine learning techniques can be used in the future for other surveys, as we plan to incorporate all possible data into our target selection.

\section*{Acknowledgements}

WAVES is a joint European-Australian project based around a spectroscopic campaign using the 4-metre Multi-Object Spectroscopic Telescope. The WAVES input catalogue is based on data taken from the European Southern Observatory’s VST and VISTA telescopes. Complementary imaging of the WAVES regions is being obtained by a number of independent survey programmes including GALEX MIS, VST, WISE, Herschel-ATLAS, and ASKAP providing UV to radio coverage. WAVES is funded by the ARC (Australia) and the participating institutions. The WAVES website is \url{https://wavesurvey.org}. Based on observations made with ESO Telescopes at the La Silla Paranal Observatory under programme ID 179.A-2004 and ID 177.A-3016. TLC is supported by the  Scientific and Technology Facilities Council (STFC), grant numbers ST/T506461/1 and ST/V507131/1. SP is also supported by the STFC, grant number ST/Y509620/1. Some parts of this work used the DiRAC Data Intensive service (CSD3) at the University of Cambridge, managed by the University of Cambridge University Information Services on behalf of the STFC DiRAC HPC Facility (\url{https://www.dirac.ac.uk}). The DiRAC component of CSD3 at Cambridge was funded by BEIS, UKRI and STFC capital funding and STFC operations grants. DiRAC is part of the UKRI Digital Research Infrastructure. ET acknowledges the ETAg grant PRG1006 and the CoE project TK202 funded by the HTM. RD gratefully acknowledges support by the ANID BASAL project FB210003. MB is supported by the Polish National Science Center through grants no. 2020/38/E/ST9/00395, 2018/30/E/ST9/00698, 2018/31/G/ST9/03388 and 2020/39/B/ST9/03494, and by the Polish Ministry of Science and Higher Education through grant DIR/WK/2018/12. DG is supported by the Deutsche Forschungsgemeinschaft (DFG, German Research Foundation) under Germany's Excellence Strategy – EXC-2094 – 390783311. AI and ML acknowledge financial support from INAF-Minigrant "4MOST- StePS: a Stellar Population Survey using 4MOST@VISTA" (2022). GK is supported by the Polish National Science Center through grant no. 2020/38/E/ST9/00395. We would also like to thank John F. Wu, Boudewijn F. Roukema, Letizia P. Cassarà, Guillaume Guiglion, Tayyaba Zafar, Alexander Fritz and Andrew Humphrey for their proofreading, comments and suggestions.

\section*{Data Availability}
The WAVES photometric input catalogue used in this work can be found under the `Input Cat Downloads' tab of \url{https://wavessegview.icrar.org}. A code tutorial of this method running on a subset of the WAVES photometric input catalogue can be found in the following GitHub repository: \url{https://github.com/toddlcook/UMAP-star-gal-separation}.



\bibliographystyle{mnras}
\bibliography{UMAP_star_gal_separation} 




\appendix

\onecolumn
\section{Visualisation of all UMAP projected dimensions}
\label{appendix:corner}
\begin{figure*}
    \includegraphics[width=\textwidth]{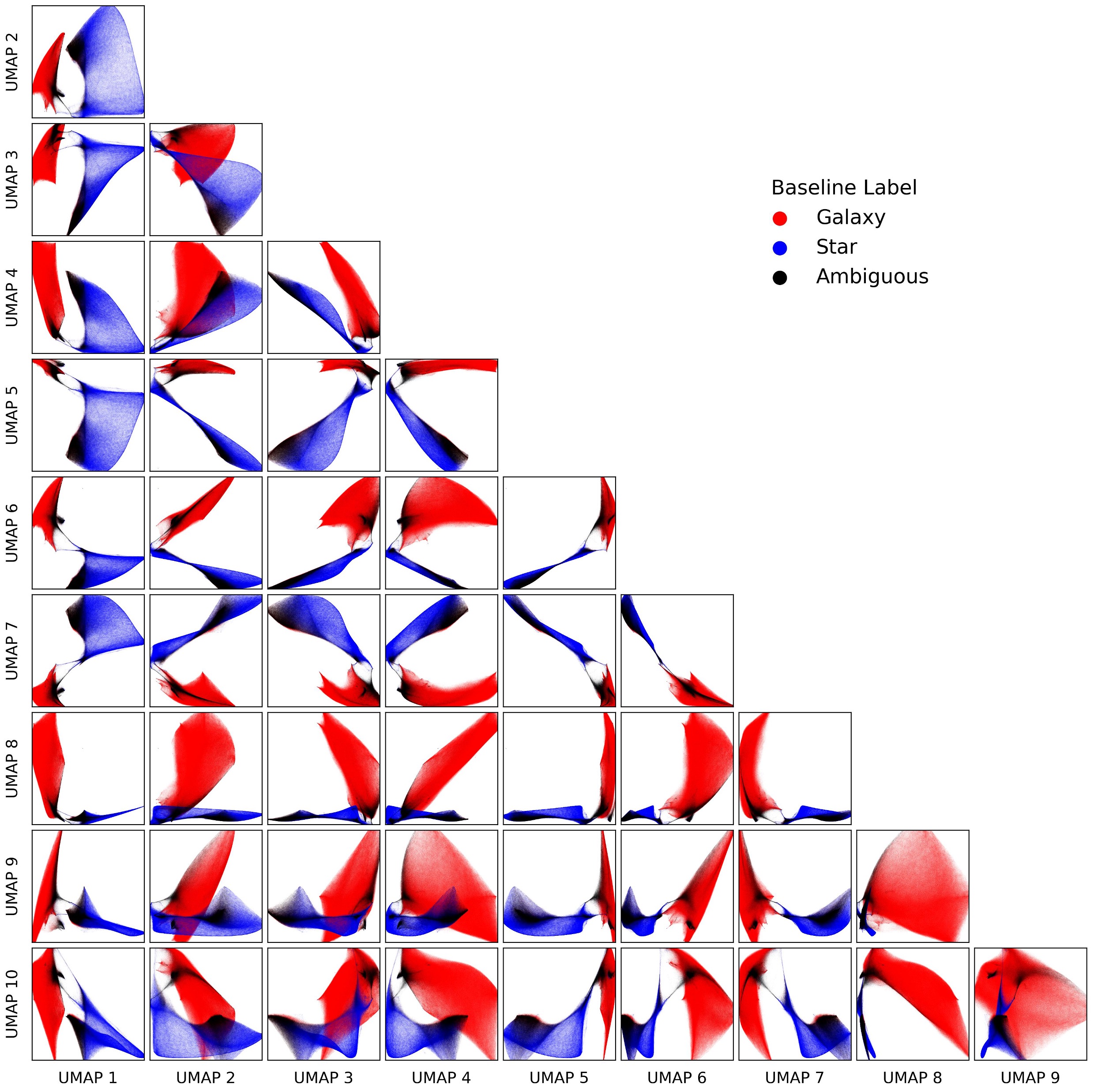}
    \caption{The same as Figure~\ref{fig:umap1_umap2_waves}, but showing the entire 10-dimensional space of UMAP projections for the WAVES-Wide South region. Colours indicate the classification made by the baseline algorithm. }
    \label{fig:UMAP_corner}
\end{figure*}

Figure~\ref{fig:UMAP_corner} shows a corner plot of the entire 10 projected dimensions of UMAP for the WAVES-Wide South region. This visualises the entire feature space which \textsc{hdbscan} is clustering. UMAP 1 versus UMAP 2 appears to show the most variance, although higher dimensions also exhibit good clustering, such as UMAP 4 versus UMAP 7 and UMAP 6 versus UMAP 10.

\newpage

\section{Source number counts}
\begin{figure*}
    \includegraphics[width=8cm]{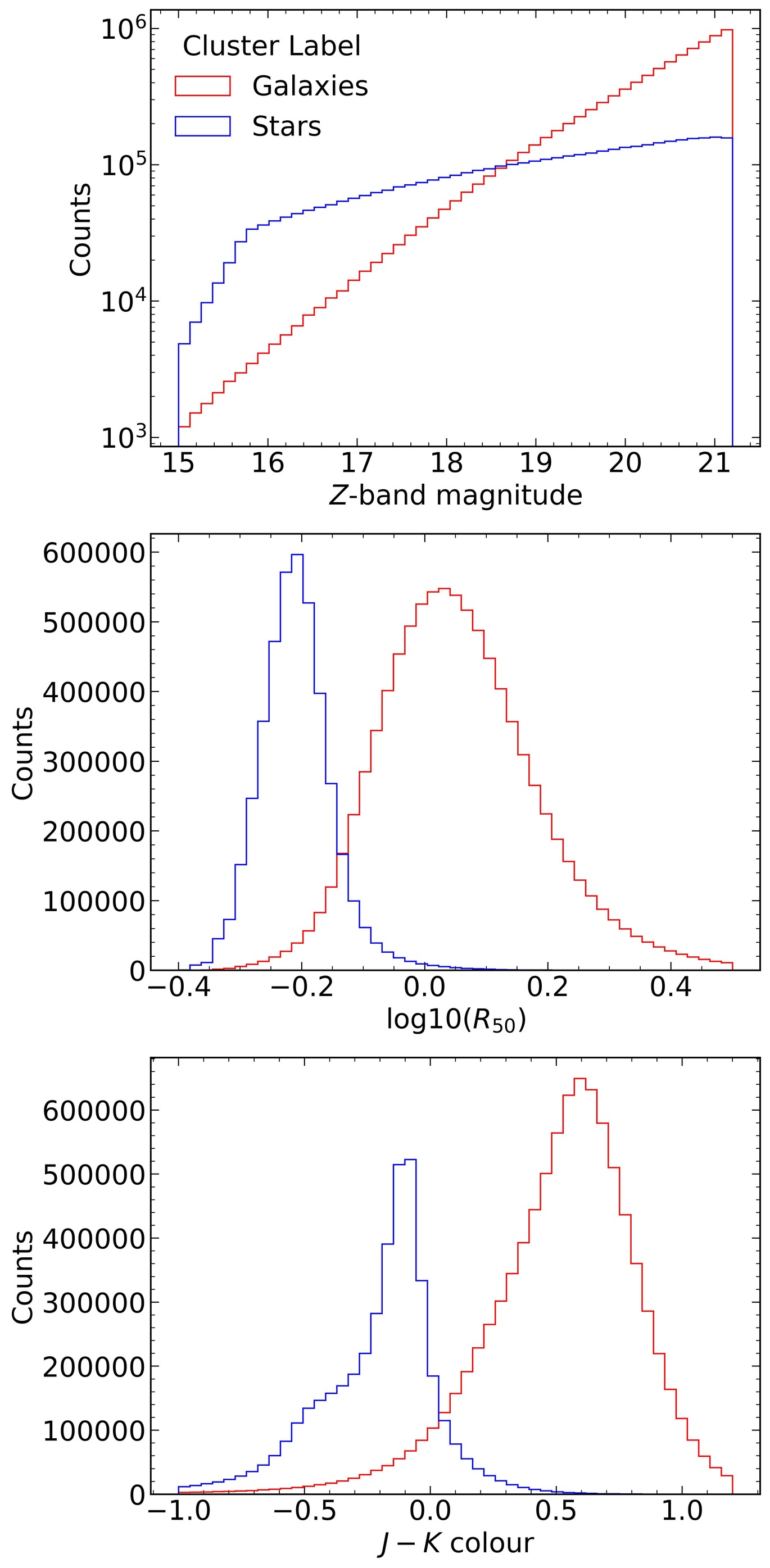}
    \caption{The number counts of our classified stars and galaxies, as a function of $Z-$band magnitude in the upper panel, the log of half-light radius in the middle panel, and $J-K_s$ colour in the lower panel.}
    \label{fig:properties}
\end{figure*}

Figure~\ref{fig:properties} shows the number counts of stars and galaxies according to our classifier as a function of brightness, radius and colour. The drop-off in the number of stars at magnitude $Z\sim16$ is due to stars being masked out brighter than a Gaia magnitude $G<16$.

\newpage

\section{WAVES imaging of exemplar `Challenging' galaxies}
\begin{figure*}
    \includegraphics[width=\textwidth]{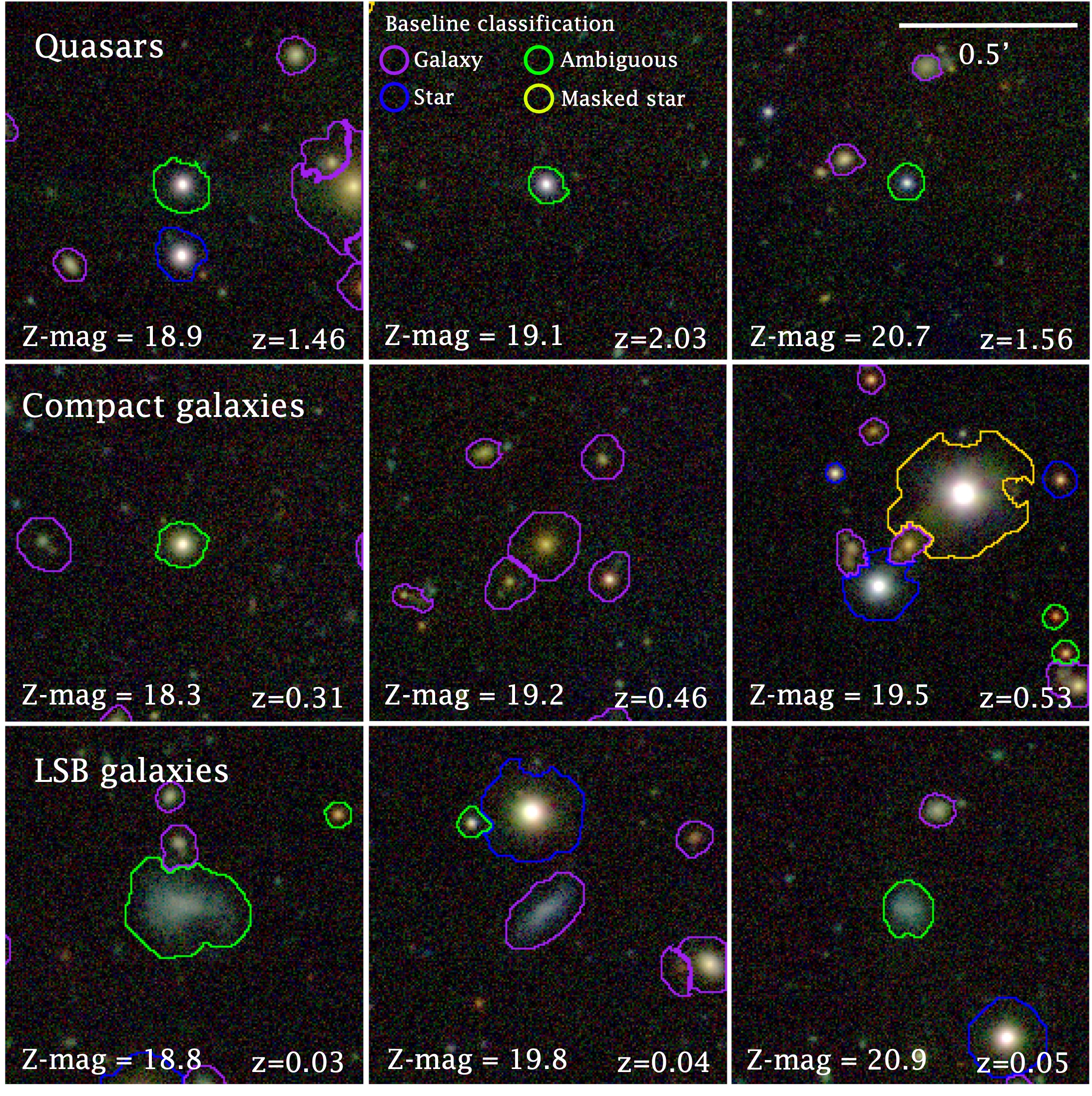}
    \caption{Nine exemplars `challenging' galaxies from the ground truth set, which we successfully classify as galaxies, taken from \url{https://wavessegview.icrar.org/}, the WAVES segment viewer. The top row are quasars, the middle row are compact galaxies and the bottom row are low surface brightness galaxies. Galaxies are fainter going from left to right. The redshifts have been obtained from spectroscopy either from GAMA or DESI. The coloured segments are produced by \textsc{ProFound}, where purple, blue and green segments indicate a baseline classification of galaxy, star and ambiguous, respectively. The yellow segments indicate a source that has been masked. Each image is 1 arcminute across. Images are compiled using KiDS $g-$ and $r$-band, and VIKING $Z$-band. }
    \label{fig:exemplar}
\end{figure*}

Figure~\ref{fig:exemplar} shows some examples of the quasars, compact galaxies and low surface brightness galaxies we classify in this work. All nine of these sources are correctly labeled by us as galaxies, but the baseline method (distinguished by the colour of the segments) is occasionally wrong. The imaging is taken from the WAVES segment viewer, \url{https://wavessegview.icrar.org/}, which can be used to view the sources identified by \textsc{ProFound} across the entire WAVES regions. It also showcases the imaging used for the input catalogue; a combination of KiDS and VIKING photometry.

\newpage

\section{Heightened clustering amplitude for discrepant classifications}
\label{tiling_section}

\begin{figure*}
    \includegraphics[width=\textwidth]{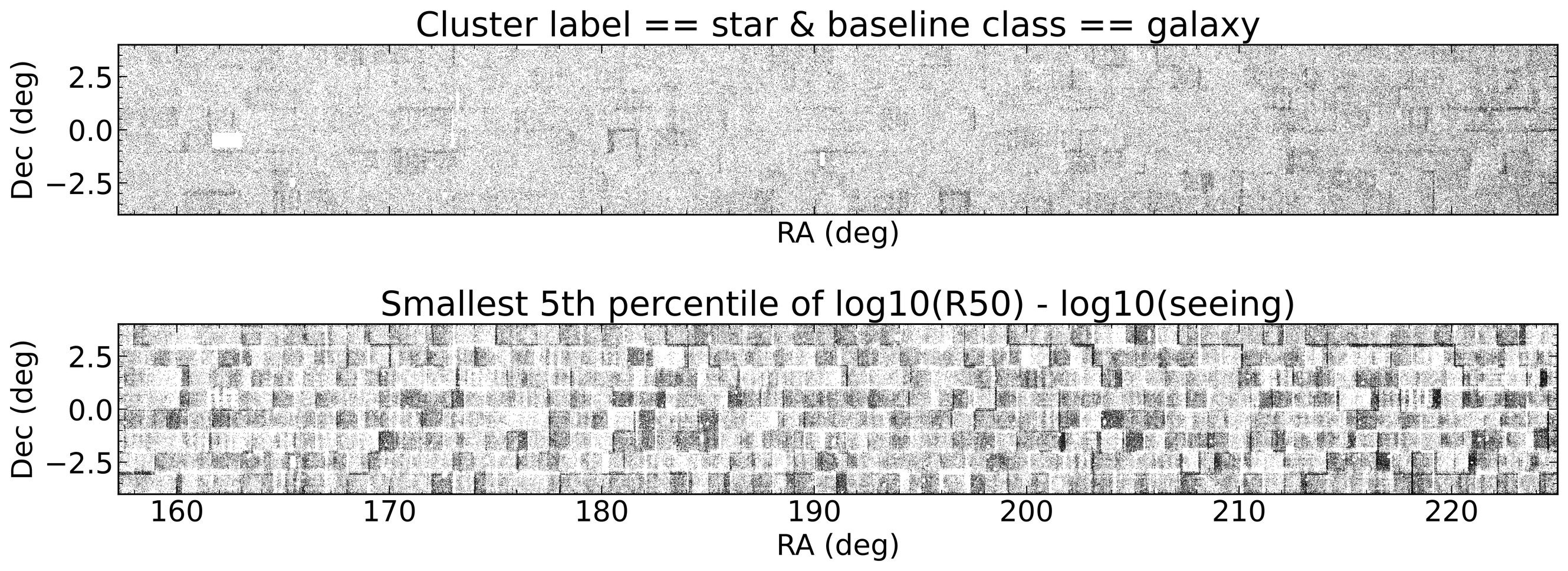}
    \caption{Upper panel: RA and declination of sources in WAVES Wide North which have been classified by our method as star and by the baseline method as a galaxy. Lower panel: sources with the smallest 5th percentile of seeing-subtracted radius, a metric that determines the baseline star-galaxy separation method. Both show evidence of systematic overdensities in square degree patterns. }
    \label{fig:tiling}
\end{figure*}

We believe the heightened amplitude in the upper right plot of Figure~\ref{fig:wtheta} (cluster galaxy, baseline ambiguous) to be due to a systematic issue within the WAVES photometric input catalogue. Plotting the sources which we label as star and the baseline method classes as galaxies (which exhibit the same heightened clustering) in the upper plot of Figure~\ref{fig:tiling}, we see a systematic tiling pattern across square degrees. This can be traced to the process in which the photometric catalogue is built, in which square degree tiles of photometry are formed, and the seeing is averaged across the tile from different observational blocks. This potentially leads to a mis-estimation of the seeing across tiles. It can be seen in the lower panel of Figure~\ref{fig:tiling} sources with the 5th smallest percentile of seeing-subtracted radius, exhibiting an exaggerated tiling pattern across the square degrees. The baseline star-galaxy separation method is dependent on the seeing-subtracted radius of each source (see Equation \ref{baseline_equation}), which is why this tiling is evident in contradictory sources, in which our method and the baseline label disagree. The non-uniform distribution of these objects leads to an increase in the amplitude of the correlations at all scales.


\bsp	
\label{lastpage}
\end{document}